\documentclass[11pt,a4paper]{article}
\pdfoutput=1
\usepackage{graphicx,rotating,hyperref,slashed,xcolor,amsmath,amssymb,amsfonts,colortbl,cite,subfigure,float,mathrsfs,mathalpha}
\usepackage{graphics}
\usepackage{comment} 
 
\usepackage[utf8]{inputenc}
\usepackage[T1]{fontenc}
\usepackage[french,english]{babel}
\usepackage{csquotes}
\usepackage{setspace}
\usepackage{geometry}
 \newcommand{\bea}{\begin{eqnarray}}
\newcommand{\eea}{\end{eqnarray}}
 
\newcommand{\be}{\begin{equation}}
\newcommand{\ee}{\end{equation}}

\hypersetup{colorlinks,bookmarksopen,bookmarksnumbered,
linkcolor=blue,pdfstartview=FitH,urlcolor=rossos,citecolor=verde}
\numberwithin{equation}{section}

 \newcommand{\sfootnote}[1]{} 
\definecolor{bluc}{cmyk}{1,0.5,0,0.1}
\definecolor{rossoCP3}{cmyk}{0,.88,.77,.40}
\definecolor{rosso}{cmyk}{0,1,1,0.4}
\definecolor{rossos}{cmyk}{0,1,1,0.55}
\definecolor{rossoc}{cmyk}{0,1,1,0.2}
\definecolor{verdes}{cmyk}{0.92,0,0.59,0.4}

\hypersetup{%
    ,urlcolor=bluc
    ,citecolor=rossos
    ,linkcolor=bluc
    }

\usepackage[utf8]{inputenc}
\usepackage{amsmath,amssymb}
\usepackage{geometry}
\usepackage{tikz}
\usepackage{enumitem}
\usetikzlibrary{arrows.meta, positioning, shapes.geometric}
\newlist{seclist}{enumerate}{3}
\setlist[seclist,1]{label=\textbf{\thesection}.\arabic*.,ref=\thesection.\arabic*}
\setlist[seclist,2]{label=\textbf{\thesection}.\arabic{seclisti}.\arabic*.,ref=\thesection.\arabic{seclisti}.\arabic*}
\setlist[seclist,3]{label=\textbf{\thesection}.\arabic{seclisti}.\arabic{seclistii}.\arabic*.,ref=\thesection.\arabic{seclisti}.\arabic{seclistii}.\arabic*}

\definecolor{bluc}{cmyk}{1,1,0,0.1}
\definecolor{rossoCP3}{cmyk}{0,.88,.77,.40}
\definecolor{rosso}{cmyk}{0,1,1,0.4}
\definecolor{rossos}{cmyk}{0,1,1,0.55}
\definecolor{rossoc}{cmyk}{0,1,1,0.2}
\definecolor{verdes}{cmyk}{0.92,0,0.59,0.4}

\begin{document}

\vspace{1truecm}
 \begin{center}
\boldmath

{\textbf{\LARGE  
{ 
Probing Gravitational-Wave Four-Point Correlators}}
}
\unboldmath
\end{center}
\unboldmath

\vspace{-0.2cm}

\begin{center}
\vspace{0.1truecm}

\renewcommand{\thefootnote}{\fnsymbol{footnote}}
\begin{center} 
{\fontsize{13}{30}
\selectfont 
Martina Ciprini \footnote{\texttt{martina.ciprini@studio.unibo.it}}, Maria Lucia Marcelli \footnote{\texttt{marialucia.marcelli@studio.unibo.it}}, Gianmassimo Tasinato \footnote{\texttt{g.tasinato2208.at.gmail.com}}
} 
\end{center}


\begin{center}

\vskip 6pt
\textsl{ Physics Department, Swansea University, SA2 8PP, UK,}
\\
\textsl{ Dipartimento di Fisica e Astronomia, Universit\`a di Bologna,\\
 INFN, Sezione di Bologna,  viale B. Pichat 6/2, 40127 Bologna,   Italy}
\vskip 4pt

\end{center}
\hskip0.1cm

\begin{abstract}
\noindent
Stochastic gravitational-wave backgrounds (SGWBs) of primordial origin
offer a powerful probe of early-Universe physics and possible
dark-sector dynamics. While most searches focus on the GW power
spectrum, additional information is encoded in higher-order correlators
that characterize the statistical properties of the signal.
In this work we study non-Gaussian features of a cosmological SGWB
generated at second order by vector fluctuations, a class of sources
well motivated in early-Universe scenarios. Within this framework we
develop  tools to characterize higher-order GW correlators and
compute representative four-point functions that generate a connected
contribution to the GW trispectrum. We show that the trispectrum
amplitude scales as the square of the GW power spectrum and peaks in
characteristic folded momentum configurations, reflecting the structure
of the nonlinear source. We then explore the observational implications.
First, we demonstrate that the connected trispectrum contributes to the
variance of two-point overlap reduction functions, including the
Hellings--Downs curve relevant for pulsar timing arrays. We then
construct the optimal estimator to measure the connected trispectrum
with ground-based interferometers. Our results highlight how
non-Gaussian SGWB statistics provide a complementary observable to
probe the origin of GW backgrounds and to distinguish cosmological from
astrophysical sources.
\end{abstract}

\end{center}

\renewcommand{\thefootnote}{\arabic{footnote}}
\setcounter{footnote}{0}

\section{
Introduction and Conclusions}

The search for a stochastic gravitational-wave background (SGWB) of primordial origin
offers a unique and powerful probe of the physics of the early Universe
\cite{Caprini:2018mtu}. Unlike electromagnetic radiation, gravitational waves
propagate essentially unimpeded from their production epoch, allowing them to
carry direct information about physical processes occurring at extremely high
energies and very early times.
A particularly compelling target consists of scenarios in which present-day
cosmic puzzles are explained through dynamics in the early Universe. Examples
include models where primordial black holes account for a fraction of the dark
matter, scenarios in which primordial vector fields seed the large-scale
magnetic fields observed in galaxies and clusters, or frameworks in which dark
matter arises from a dark photon sector. See, for example,
\cite{Ozsoy:2023ryl,Durrer:2013pga,Fabbrichesi:2020wbt} for reviews of these
possibilities.
These scenarios typically involve the amplification of primordial fluctuations
or the dynamics of additional fields in the early Universe. As a generic
consequence of their non-linear evolution, they inevitably source gravitational
waves at second order in perturbations, leading to the production of a
stochastic gravitational-wave background. In this sense, the SGWB constitutes
an unavoidable byproduct of the mechanisms responsible for generating the
underlying cosmological structures or dark-sector components. See, e.g.,
\cite{Domenech:2021ztg,LISACosmologyWorkingGroup:2025vdz} and references
therein.
Importantly, the resulting SGWB signals are expected to display distinctive
spectral and statistical properties that encode information about their
production mechanisms. Characterizing these features therefore provides a
powerful avenue to discriminate between different early-Universe scenarios and
to probe the physics of the dark sector.

A central challenge in the context of SGWB physics    is the identification of observables capable
of discriminating among different gravitational-wave sources, whether primordial \cite{Caprini:2018mtu}
or astrophysical \cite{Regimbau:2011rp}, and of accurately extracting their physical
properties.
Several strategies have been proposed to distinguish cosmological from
astrophysical SGWB, including the study of their spectral profiles (see e.g.
\cite{Caprini:2018mtu,Kuroyanagi:2018csn,Caprini:2019pxz,LISACosmologyWorkingGroup:2022jok} and references therein) and of their anisotropies: see e.g. \cite{Contaldi:2016koz,Bartolo:2019oiq,Bartolo:2019yeu,LISACosmologyWorkingGroup:2022kbp}.

In this work we investigate primordial non-Gaussian features in the
distribution of gravitational waves as potential observables for
discriminating among different sources of the SGWB. Non-Gaussianity has long been
recognized as a powerful probe of early-Universe physics. In
particular, primordial non-Gaussianities generated during inflation
have been extensively studied in the scalar sector of curvature
perturbations (see e.g.~\cite{Bartolo:2004if}), and constraints of their amplitude
and shape in the cosmic microwave background have played a crucial
role in ruling out classes of inflationary models
\cite{Planck:2013wtn,Planck:2019kim}.
In comparison, much less attention has been devoted to the study of
primordial non-Gaussianity in SGWBs of cosmological origin, and to the
development of suitable estimators to detect such signals with
gravitational-wave experiments. Our work aims to contribute to this
direction.
We focus on the class of dark-sector scenarios discussed above, in
which gravitational waves are generated at second order in
fluctuations. Since the GW signal arises from non-linear sources, the
resulting SGWB is expected to exhibit intrinsically non-Gaussian
statistics. This property provides an additional observable that can
be exploited to discriminate among different GW production
mechanisms.
In this work we therefore investigate primordial SGWB non-Gaussianity
generated at second order in dark-sector models, and assess its
potential as a diagnostic of the underlying early-Universe dynamics.
We show how non-Gaussian features encode information about the
physical processes responsible for GW production, and we discuss the
prospects for detecting such signatures with present and future
gravitational-wave experiments. Our analysis aims to bridge a gap
between theoretical predictions and observational strategies,
exploring small-scale SGWB non-Gaussianities by working at  the interface between
gravitational-wave theory and experimental searches.

Astrophysical SGWBs, being generated by the cumulative contribution of
a large number of independent sources, are expected to be nearly
Gaussian by virtue of the central limit theorem
\cite{Allen:1996vm,Adshead:2009bz}. This expectation holds provided
that a sufficiently large population of sources contributes to the
signal.
An important exception arises when the SGWB is produced by a limited
number of unresolved sources. In this regime the signal can exhibit a
so-called ``popcorn'' character, reflecting the Poissonian statistics
of the underlying astrophysical events. The resulting background is
then characterized by a distinct form of non-Gaussianity associated
with the discreteness of the sources. The search for and
characterization of such signals is an active area of research; see,
for example,
\cite{Drasco:2002yd,Allen:2002jw,Seto:2008xr,Seto:2009ju,Adshead:2009bz,Zhu:2011bd,Martellini:2014xia,Cornish:2015pda,Buscicchio:2022raf,Ballelli:2022bli,Lawrence:2023buo,Falxa:2025qxr}.
See also \cite{Romano:2016dpx,Renzini:2022alw} for  comprehensive reviews. 

Cosmological gravitational-wave backgrounds, by contrast, originate
from processes operating in the early Universe and can exhibit
coherence over very large scales. As a consequence, they may possess
non-vanishing higher-order correlators with statistical properties
that differ qualitatively from the Poisson-type non-Gaussianities
expected in astrophysical backgrounds.
Non-Gaussian features in the SGWB therefore constitute a potentially
powerful observable for distinguishing among different classes of GW
sources.
 Symmetry arguments and specific model-building approaches
 are known to 
 provide powerful constraints on the possible shapes and properties of cosmological
GW  non-Gaussianities (see, e.g., \cite{Maldacena:2011nz,Soda:2011am,Shiraishi:2011st,Agrawal:2017awz,Gao:2011vs,Bordin:2016ruc,Bartolo:2018qqn,Ozsoy:2019slf,DeLuca:2024asq}). Here 
we show that features of a family of induced GW sources allow
us to accurately characterize  the structure of observable non-Gaussian GW correlators, and to investigate their specific
observational consequences. 

We focus on SGWBs induced at second order by vector fluctuations, see e.g.~\cite{Durrer:1999bk,Caprini:2001nb,Mack:2001gc,Shaw:2009nf,Saga:2018ont,Ozsoy:2023gnl,Bhaumik:2025kuj,Maiti:2025cbi,Atkins:2025pvg}. These constitute the vector-field analogue of the well-studied case of scalar-induced SGWBs, extensively investigated in primordial black hole scenarios, starting from~\cite{Matarrese:1993zf,Matarrese:1997ay,Ananda:2006af,Baumann:2007zm,Saito:2009jt,Espinosa:2018eve,Kohri:2018awv}. This setup, described in 
Section \ref{sec_setup}, provides a useful benchmark scenario, as it leads to technically simpler and physically transparent calculations of GW correlators. In particular, time and momentum integrations factorize within the relevant convolution integrals, allowing the time integrals to be performed analytically. This procedure isolates characteristic non-Gaussian quadrilateral shapes for
the Fourier momenta, 
associated with \emph{folded} configurations,  and they naturally leads to the notion of \emph{stationary non-Gaussianity}~\cite{Powell:2019kid,Tasinato:2022xyq}.  
We extend this framework by carrying out in Section \ref{sec_trispectrum} an analytic study of representative GW four-point functions, which generate a connected contribution to the GW trispectrum. (We do not consider spin-2 GW three-point functions, since they are not detectable with pulsar timing arrays or ground-based interferometers, see e.g.~\cite{JimenezCruz:2025wqa}.) We show that the amplitude of the GW trispectrum scales as the square of the GW power spectrum. 

In
Section \ref{sec_det}
we then investigate the observational implications of these results, demonstrating that the connected trispectrum can contribute to the variance of two-point overlap reduction functions -- most notably to the variance of the Hellings--Downs curve relevant for pulsar timing array experiments. Finally, we analytically compute the four-point overlap reduction functions for ground-based interferometers and construct the corresponding optimal estimator for their detection. 

We hope that our findings will contribute to advancing the theoretical
understanding of gravitational-wave non-Gaussianities in well-motivated
early-Universe scenarios, while also helping to identify robust
observational signatures that could be probed by current and future
gravitational-wave experiments.

\section{
Our set-up, and the SGWB
two-point function}
\label{sec_setup}

We study gravitational waves (GW) produced  during the era of radiation
domination, induced at second order in perturbations 
 by primordial magnetic field fluctuations generated during cosmic inflation. Such magnetic fields can be associated
 with Standard Model electromagnetism and primordial magnetogesis \cite{Durrer:2013pga}, or they can represent the magnetic-field components
 of dark vector models aimed at describing dark matter 
 via longitudinal massive vector dynamics 
 (see e.g. \cite{Graham:2015rva}). In our work, we consider a vector-induced 
 SGWB as 
benchmark scenario 
 for the broad family of models able to produce a background of GW at second order in fluctuations. 
 In this and the next
 section we compute two-point and non-Gaussian
 correlators respectively. We show that our set-up  offers
  advantages with respect to the scalar-induced case, leading to technically feasible and physically
  transparent calculations.
 Then, in Section \ref{sec_det},  
 we   analyse how GW non-Gaussianities 
   can
 offer new handles to probe or constrain properties
 of dark sector physics 
 with GW experiments.

\paragraph{An example of vector
field models from the early universe.} Scenarios involving
vector fields produced in the primordial universe arise
in 
different contexts. To
consider a
concrete example, one of  the simplest
models of  magnetogenesis from cosmic inflation  is based
on the Ratra action for the vector sector \cite{Ratra:1991bn}
\be
\label{eq_ratra}
S\,=\,-\frac14\,\int d^4 x\,\sqrt{-g}\,
f(\phi)\,F_{\mu\nu}F^{\mu\nu}\,,
\ee
where $F_{\mu\nu}=\partial_\mu A_\nu-\partial_\nu A_\mu$
is the vector field strength, and $f(\phi)$
a function of the inflaton field and its gradient, which 
characterise the dynamics of the    vector. We work in the mostly-plus metric convention. We assume a Friedmann--Robertson--Walker background metric with scale factor $a(\tau)$, with $\tau$ denoting the conformal time. We focus on the magnetic
part of the vector spectrum only. We decompose
the magnetic field in Fourier modes as 
\bea
\label{eq_fou}
B_i(\tau,\mathbf{x})
&=&
\int \frac{d^3\mathbf{k}}{(2\pi)^3}\,
e^{i\mathbf{k}\cdot\mathbf{x}}\, B_i(\tau,\mathbf{k})
\nonumber\\
&=&
\sum_{\lambda}\int \frac{d^3\mathbf{k}}{(2\pi)^3}\,
e^{i\mathbf{k}\cdot\mathbf{x}}\,
e^{(\lambda)}_i(\hat{k})\, B_{\mathbf{k}}(\tau),
\eea 
The spin-one polarization tensors $e^{(\lambda)}_i(\hat{k})$
are defined in circular basis 
with helicity $\lambda=R/L=\pm1$. 
They are constructed from two orthonormal unit vectors $\hat p$ and $\hat q$, transverse to the direction $\hat k$ 
\be
\label{def_vpt}
e^{(\lambda)}_i(\hat{k})
=
\frac{\hat p_i + i\,\lambda\,\hat q_i}{\sqrt{2}}\,.
\ee
Hatted vectors denote unit vectors, e.g.\ $\hat{k}=\mathbf{k}/k$ with $k=|\mathbf{k}|$.  
We assume that inflationary dynamics generate a magnetic-field spectrum  characterized by the following
two-point function in Fourier space
\begin{eqnarray}
\label{eq_mfa}
\langle B_i(\tau,\mathbf{k})\, B^*_j(\tau,\mathbf{q}) \rangle
&=&
(2\pi)^3\,\delta({\bf k} -{\bf q})
\,\langle B_i(\tau,\mathbf{k})\, B^*_j(\tau,\mathbf{k}) \rangle_\Delta\,,
\label{def_smfs}
\end{eqnarray}
where
\begin{eqnarray}
\langle B_i(\tau,\mathbf{k})\, B^*_j(\tau,{\mathbf{k}}) \rangle_\Delta
&=&
\pi_{ij}(\hat{k})\, P_B(k)\,. 
\end{eqnarray}
The Dirac-delta in  Fourier momenta appearing in Eq.~\eqref{def_smfs} is associated
with the rotational invariance  of the background. 
The index $\Delta$ in $\langle \dots  \rangle_\Delta$  denotes correlators factorising the aforementioned delta function. 
The tensor $\pi_{ij}$ is a combination of spin-1 polarization
tensors
\be
\pi_{ij}(\hat{k})=
\sum_{\lambda=\pm 1} e_i^{(\lambda)}(\hat{k})
e_j^{(\lambda) *}(\hat{k})\,=\,
\delta_{ij}-\hat{k}_i\hat{k}_j,
\ee
For convenience  we define the rescaled magnetic field spectrum $\mathcal{P}_B(k)$ through the formula
\be
\label{ps_resc}
P_B(k)\equiv \langle B_{\bf k}(\tau)B^*_{\bf k}(\tau) \rangle_{\Delta} \,=\, \frac{2\pi^2}{k^3}\,\mathcal{P}_B(k)\,.
\ee
The explicit scale dependence of the function $\mathcal{P}_B(k)$
depends on the set-up one considers -- the model of inflation, and
the structure of the function $f$ in the specific model of Eq.~\eqref{eq_ratra}. See e.g. \cite{Durrer:2013pga} for a review.
Besides couplings as Eq.~\eqref{eq_ratra}, one can also consider models of massive
vectors during inflation, with  interesting consequences for the dark
matter problem \cite{Graham:2015rva}. In all these cases the magnetic field spectrum
can get enhanced at small scales, and our considerations in principle apply to all these scenarios.
We shall assume though that vector perturbations
propagate around a system with a vanishing vector field background, 
 so that spin-1 fluctuations have Gaussian statistics -- although, as we shall learn, they can lead to non-Gaussianities in the induced spin-2, GW background. 

 \paragraph{Computation of the induced SGWB spectrum} An enhanced
 magnetic field spectrum ${\cal P}_B(k)$ at small scales can source a background of 
 GW at second order in fluctuations, after inflation ends. 
 In this Section we focus on two-point correlation functions
 and compute the induced GW spectrum, while in  Section~\ref{sec_trispectrum}
 we discuss SGWB non-Gaussianities. 
 
 We decompose tensor perturbations corresponding to GW as
\be
\label{fou_dec}
h_{ij}(\tau,\mathbf{x})
=
\sum_{\lambda}
\int \frac{d^3\mathbf{k}}{(2\pi)^3}\,
e^{i \mathbf{k}\cdot\mathbf{x}}\,
e^{(\lambda)}_{ij}(\hat{{k}})\,
h^{(\lambda)}_{\mathbf{k}}(\tau),
\ee
where $e^{(\lambda)}_{ij}(\hat{{k}})$ are the
spin-2, transverse--traceless (TT) polarization tensors
in circular basis $(R,L)=\pm2$~\footnote{The
spin-2 polarization tensors
in circular basis $(R,L)=(+2,-2)$ are related to the ones in $(+,\times)$ basis
by the formula
\be
\label{rel_cicr}
e^{(\pm 2)}_{ij}(\hat k)\,=\,\frac{e^{(+)}_{ij}(\hat k) \pm i\, e^{(\times)}_{ij}(\hat k)}{\sqrt{2}}\,.
\ee
We assume that the polarization tensors $e^{(+,\times)}_{ij}$
are real, hence $\left(e^{(\pm 2)}_{ij}\right)^* = e^{(\mp 2)}_{ij}$.}.
The spin-2 polarization tensors are built as product
of spin-1 ones of Eq.~\eqref{def_vpt}:
\be
\label{sqpolte}
e^{(\lambda)}_{ij}(\hat{{k}})\,=\,
e^{(\lambda)}_{i}(\hat{{k}}) e^{(\lambda)}_{j}(\hat{{k}})\,.
\ee
The following combination of spin-2 polarization tensors
defines the projector operator, an important quantity in our analysis:
\begin{eqnarray}
\Lambda_{ij \ell m}
&=& \sum_{\lambda=\pm2}\,e_{ij}^{(\lambda)} e^{(\lambda)*}
_{\,\ell m}
\nonumber
\\
&=&e_{ij}^{(L)} e^{(L)*}
_{\,\ell m}+e_{ij}^{(R)} e^{(R)*}
_{\,\ell m}
=\,e_{ij}^{(+)} e^{(+)}_{\,\ell m}+e_{ij}^{(\times)} e^{(\times)}_{\,\ell m}
\nonumber
\\
\label{def_ofLA}
&=&
\frac12\Big(
\pi_{i \ell} \pi_{j m}
+\pi_{j \ell} \pi_{i m}
-\pi_{ij}\pi_{\ell m}
\Big).
\end{eqnarray}
which projects a two-index tensor in its transverse-traceless
components.

In order to ensure that $h_{ij}(\tau, {\bf x})$
is real, we assume
 $h^{(\lambda)*}_{\mathbf{k}}(\tau) = h^{(-\lambda)}_{-\mathbf{k}}(\tau)$. 
The two-point correlators for
Fourier components satisfy the condition 
 \be
\langle
h_{\mathbf{k}}^{(\lambda_1)}(\tau)\,
h_{\mathbf{q}}^{(\lambda_2)*}(\tau)
\rangle
 = {(2\pi)^3}\delta^{(3)}({\bf k}- {\bf q})
 \,\delta_{\lambda_1,\lambda_2}\,
 \langle
h_{\mathbf{k}}^{(\lambda_1)}(\tau)\,
h_{\mathbf{k}}^{(\lambda_1)*}(\tau)
\rangle_\Delta\,.
 \ee
 The role of the $\delta$-function over momenta
 has been discussed earlier. Here, the Kronecker $\delta$-function
in the polarization indexes is related to
the conservation of helicity for the spin-2 massless modes~\footnote{This is an important point on which we return again in  Section~\ref{sec_trispectrum} when discussing 
non-Gaussian correlators.}. 
Using the previous Fourier decomposition,
 we
define
the tensor power spectrum  as
\be
\label{def_tenps}
{\cal P}_{h}(k)\,=\,\frac{k^3}{4\pi^2}
\sum_\lambda\,
\langle
h_{\mathbf{k}}^{(\lambda)}(\tau)\,
h_{\mathbf{k}}^{(\lambda)*}(\tau)
\rangle_\Delta\,.
\ee 
This  is the quantity we wish to compute
in this Section, reviewing and in part extending results
developed e.g. in \cite{Atkins:2025pvg,Ragavendra:2026fgs}.

\bigskip

Given an early universe model producing 
 magnetic fields, 
our computation of the second order,  induced GW spectrum follows \cite{Atkins:2025pvg}. 
 After inflation ends, for $\tau\ge \tau_f$, the GW Fourier mode functions  obey
\be
\label{eq_eveqh}
h_{\mathbf{k}}^{(\lambda)\,\prime\prime}(\tau)
+2\mathcal{H}\,h_{\mathbf{k}}^{(\lambda)\,\prime}(\tau)
+k^2 h_{\mathbf{k}}^{(\lambda)}(\tau)
=
S^{(\lambda)}(\tau,\mathbf{k}),
\ee
where primes indicate derivative along conformal time, ${\cal H}=a'/a$, and  the source term on the
right-hand-side of this equation is induced by the anisotropic stress
 associated with magnetic
fields. It  reads
\bea
\label{eq_gwsor}
S^{(\lambda)}(\tau,\mathbf{k})
&=&
\frac{2}{a^2(\tau)}\,
e^{(\lambda)\,ij}(\hat{{k}})\,
\Lambda_{ij}{}^{mn}(\hat{{k}})\,
\tau^{(B)}_{mn}(\mathbf{k}) 
\nonumber
\\
&=&\frac{2}{a^2(\tau)}\,
e^{(\lambda)\,ij}(\hat{{k}})\,
\tau^{(B)}_{ij}(\mathbf{k}) 
.
\eea
The magnetic-field stress tensor in Eq.\eqref{eq_gwsor} is given by (see e.g. \cite{Mack:2001gc})
\be
\label{eq_emtmf}
\tau^{(B)}_{ij}(\mathbf{k})
=
\frac{1}{4\pi}
\int \frac{d^3\mathbf{p}}{(2\pi)^3}
\left[
B_i(\mathbf{p}) B_j(\mathbf{k}-\mathbf{p})
-\frac{\delta_{ij}}{2}
B_m(\mathbf{p}) B_m(\mathbf{k}-\mathbf{p})
\right].
\ee
Since $\Lambda_{ii}{}^{\ell m}=\Lambda_{ij}{}^{\ell\ell}=0$, the  term 
proportional to $\delta_{ij}$ in the previous expression does not contribute once contracted
with the projector $\Lambda$. 
The formal solution of Eq.~\eqref{eq_eveqh} can be written as
\be
\label{eq_frh}
h_{\mathbf{k}}^{(\lambda)}(\tau)
=
\frac{1}{a(\tau)}
\int d\tau'\,
g_k(\tau,\tau')\,
a(\tau')\,S^{(\lambda)}(\tau', \mathbf{k}),
\ee
where during radiation domination (RD) the Green function is
\be
g_k(\tau,\tau')
=
\frac{1}{k}
\,
\sin\left[ k \left(\tau-\tau'\right)\right]
\,.
\ee
As in \cite{Caprini:2001nb},
  we parameterize the scale factor at late times  during  this epoch as
\be
\label{scf_rd}
a(\tau)=H_0\sqrt{\Omega_{\rm rd}}\,\tau,
\ee
with $H_0$ the present-day Hubble parameter, and $\Omega_{\rm rd}=\frac{\rho_{\rm rad}}{\rho_{\rm cr}}$
the radiation density parameter.
The tensor power spectrum  defined in Eq.~\eqref{def_tenps} results
\bea
\label{eq_ts2s}
\mathcal{P}_h(k)
= 
\frac{k^3}{4\pi^2 a^2(\tau)}
\!\int\! d\tau_1 d\tau_2\,
g_k(\tau,\tau_1) g_k(\tau,\tau_2)\,
a(\tau_1)a(\tau_2)
\left(\sum_\lambda  \langle
S^{(\lambda)}(\tau_1, \mathbf{k})
S^{(\lambda)*}(\tau_2, \mathbf{k})
\rangle_\Delta \right).
\nonumber\\
\eea
Using Eq.~\eqref{eq_emtmf} and Wick theorem, the source two-point correlator 
summed over helicities becomes

\bea
{
\sum_{\lambda}\langle S^{(\lambda)}_{\bf k}(\tau_1)
    S^{(\lambda)\,*}_{\bf k}(\tau_2)\rangle_{\Delta}} &=&
\frac{4}{(4\pi)^2 a^2(\tau_1)a^2(\tau_2)}
\int d^3p\,
\Lambda_{ij}{}^{\ell n}(\hat{\mathbf{k}})
\big[
\pi^i{}_\ell(\hat{ {\bf p}})
\pi^j{}_n(\hat{ {\bf n}})
+\pi^i{}_n(\hat{ {\bf p}})
\pi^j{}_\ell(\hat{ {\bf n}})
\big]
\nonumber\\
&&\times
P_B(p)\,
P_B(|\mathbf{k}-\mathbf{p}|)\,,
\eea
where $\hat{ {n}}=(\mathbf{k}-\mathbf{p})/|\mathbf{k}-\mathbf{p}|$,
and we used the definition \eqref{eq_mfa} for the magnetic field power spectrum.
Non-Gaussian connected contributions are neglected, since as explained
above we consider set-up of Gaussian vector fields only.
Following \cite{Baumann:2007zm,Kohri:2018awv}, we introduce
the variables
\be
u=\frac{|\mathbf{k}-\mathbf{p}|}{k},
\qquad
v=\frac{p}{k},
\qquad
\mu=\frac{\mathbf{k}\!\cdot\!\mathbf{p}}{kp}
=\frac{1-u^2+v^2}{2v}.
\ee
Contracting the tensor structure above (see e.g. \cite{Caprini:2003vc}) one finds
\bea
\mathcal{C}_0(u,v)
&=&\Lambda_{ij}{}^{\ell n}(\hat{\mathbf{k}})
\big[
\pi^i{}_\ell(\hat{ {p}})
\pi^j{}_n(\hat{ {n}})
+\pi^i{}_n(\hat{ {p}})
\pi^j{}_\ell(\hat{ {n}})
\big]
\nonumber
\\
&=&
(1+\mu^2)
\left(
1+\frac{(1-\mu v)^2}{u^2}
\right).
\eea
Using $d^3p=2\pi k^3 v^2\,dv\,d\mu$, we obtain -- in terms of the rescaled
magnetic field spectrum ${\cal P}_B$ 
of Eq.~\eqref{ps_resc}:
\bea
\sum_\lambda
\langle
S^{(\lambda)}_{\mathbf{k}}(\tau_1)
S^{(\lambda)*}_{\mathbf{k}}(\tau_2)
\rangle_\Delta
&=&
\frac{{2} 
\pi^3}{k^3 a^2(\tau_1)a^2(\tau_2)}
\int_0^\infty\!\frac{dv}{u^3 v}
\int_{-1}^1\!d\mu\,
\mathcal{P}_B(ku)\,
\mathcal{P}_B(kv)\,
\mathcal{C}_0(u,v)\,.
\nonumber\\
\label{eq_sigV2}
\eea
Substituting Eq.~\eqref{eq_sigV2} into Eq.~\eqref{eq_ts2s}, the tensor spectrum ${\cal P}_h$
defined in Eq.~\eqref{eq_ts2s}  factorizes in two pieces as
\be
\label{exp_ph2}
\mathcal{P}_h(\tau,k)
=
\frac{\pi}{{2}a^2(\tau)}\,
\mathcal{I}_\tau^2\,
\mathcal{I}_{B},
\ee
with
\bea
\label{eq_tin}
\mathcal{I}_\tau^2
&=&
\left(
\int_{{\tau_R}}^{\tau} d\tau_1\,
\frac{g_k(\tau, \tau_i)}{a(\tau_1)}
\right)^2,
\\
\label{eq_min}
\mathcal{I}_{B}
&=&
\int_0^\infty \frac{dv}{u^3 v}
\int_{-1}^1 d\mu\,
\mathcal{P}_B(ku)\mathcal{P}_B(kv)
\mathcal{C}_0(u,v).
\eea

\noindent
It is important to emphasize that the expression in  Eq.~\eqref{exp_ph2} exhibits a \emph{factorizable} structure in its time and momentum integrations. As a consequence, the time integral is easily tractable and can be carried out analytically. This property will play a central role in our analysis of higher-order correlators in the next section. It considerably streamlines the computations and yields physically transparent results, particularly in clarifying which momentum configurations and non-Gaussian
shapes are permitted for $n$-point functions within the setup under consideration.

\smallskip

The time integral \eqref{eq_tin} is
independent from the magnetic field source --
it controls the effects of propagation during RD.
Instead, the momentum integral $\mathcal{I}_{B}$
of Eq.~\eqref{eq_min} 
is time independent during RD  -- while it depends
on the primordial  magnetic field spectrum ${\cal P}_B$. 
 During radiation domination, $a(\tau)/a(\tau_1)=\tau/\tau_1$ and $aH=1/\tau$:
see Eq. \eqref{scf_rd}. Averaging over rapid oscillations, and taking the large $\tau$ limit within radiation domination, yields for the time integral
\be
\label{val_itau}
\bar{\mathcal{I}}_\tau^2
=
\frac{1}{2k^2 H_0^2\,\Omega_{R}}
\left[
\mathrm{Ci}^2(| k {\tau_R}|)
+
\Big(\tfrac{\pi}{2}-\mathrm{Si}(| k {\tau_R} |)\Big)^2
\right].
\ee
with $\mathrm{Ci}$ and $\mathrm{Si}$
respectively the cosine and sine integral functions, and the bar over  $\bar{\mathcal{I}}$ indicates
average over rapid oscillations. We used the fact that in RD $ a^2 H= H_0 \sqrt{\Omega_R}$. Hence  the time integral ${\cal I}_\tau^2$ becomes
independent from time for $\tau$ sufficiently large in RD. 
The quantity
within square parenthesis in Eq.~\eqref{val_itau}
has a mild, logarithmic dependence
on momenta in the limit of small $|k \tau_f|$. See also \cite{Atkins:2025pvg}
and references therein. 
{Having introduced the averaged time integral, we define the averaged power spectrum as
\begin{equation}
    \bar{\mathcal{P}}_h\equiv\frac{1}{2}\frac{\pi}{2a^2(\tau)}\bar{{\cal I}}_\tau^2\,{\cal I}_{B}
    \label{ph_bar}\,.
\end{equation}}

We plug \eqref{val_itau} in Eq.~\eqref{exp_ph2}, and evaluate
the scale factor at time $\tau$ when the mode of momentum 
$k$ crosses the horizon during RD: $a^2(\tau_k)= \Omega_{R} H_0^2/k^2$, finding
\be
\label{exp_ph3}
\mathcal{{\bar{P}}}_h(\tau_k,k)
=
\frac{\pi}{{8}
\Omega_{R}^2 H_0^4}\,
\left[
\mathrm{Ci}^2(| k\tau_R |)
+
\Big(\tfrac{\pi}{2}-\mathrm{Si}(| k\tau_R |)\Big)^2
\right]\,
\mathcal{I}_{B}\,.
\ee
In order to take into account of the GW redshift from
horizon crossing to today, we  include effects of  transfer function
 to the previous formulas, amounting to the inclusion of an overall factor of 
$a^2(\tau_k)/2$ (see e.g. \cite{Maggiore:2018sht}), hence today we have
\be
\label{exp_ph3}
\mathcal{{\bar{P}}}_h(k)
=
\frac{\pi}{{16}
\Omega_{R} H_0^2\,k^2}\,
\left[
\mathrm{Ci}^2(| k\tau_R |)
+
\Big(\tfrac{\pi}{2}-\mathrm{Si}(| k\tau_R |)\Big)^2
\right]\,
\mathcal{I}_{B}\,.
\ee

\paragraph{Explicit  examples.} 
We are now left with the computation of $\mathcal{I}_{B}$
in Eq.\eqref{eq_min} 
--
which is the quantity that depends on ${\cal P}_B$.
 We {\it do not} limit our
attention to power-law profiles for the vector spectrum ${\cal P}_B$, as often considered, and we
instead allow for richer momentum dependences,  motivated
by recent advances in inflationary model buildings including non-slow-roll phases (see
e.g. \cite{Atkins:2025pvg,Ragavendra:2026fgs} for 
models of primordial magnetogenesis including the effects
of such phases). We keep the discussion broad
without focussing on explicit model building. Instead we consider general, simple templates with a peak
in the magnetic field spectrum.

As a  first example, 
 we focus on  the case of delta-like magnetic
spectrum, which can be motivated by models of inflation with ultra-slow-roll phases able to produce
sharp peaks in the spectrum
\be
\label{cond_delsp}
\mathcal{P}_B(k)=A_B\,\delta\left(\ln\left[k/k_\star\right] \right)\,,
\ee
with $k_\star$ a characteristic scale indicating the epoch during inflation when departures
from slowroll occur.
 A scenario with such sharp peaks is not easy to realize in practice, but
 this set-up has the advantage of being analytically tractable.
 In fact,
changing variables
as $u=(t+s+1)/2$, $v=(t-s+1)/2$, the integral ${\cal I}_{B}$
containing delta-functions 
can be explicitly evaluated, giving
\bea
{\mathcal{I}}_{B}(k)
&=&
\int_0^\infty\!dt
\int_{-1}^1\!ds\,
(1-s+t)^{-2}(1+s+t)^{-2}\,
\mathcal{P}_B(k u)\,
\mathcal{P}_B(k v)\,
\mathcal{C}_0(t,s).
\label{fin_inta}
\\
&=&
A_B^2\,\dfrac{k^2(k^2+4 k_\star^2)^2}{128\,k_\star^6}\,\Theta(2 k_\star-k)\,.
\eea
\begin{figure}[t!]
    \centering
    \includegraphics[width=0.5\linewidth]{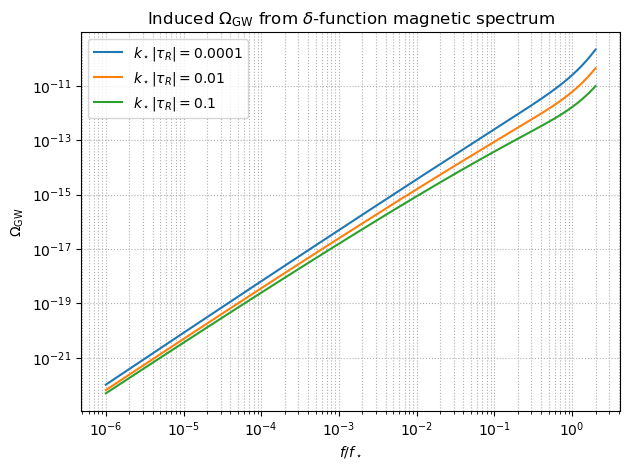}
    \caption{\small The GW energy spectrum $\Omega_{\rm GW}$ associated a monochromatic (delta-function) magnetic power spectrum, plotted as a function of $f/f_\star$ for different values of $k_\star|\tau_R|$. See
    Eq.~\eqref{Omega_deltafunction}. Notice that the the spectrum drops to zero for $f/f_\star\ge2$.  }
    \label{delta}
\end{figure}
Hence the final expression for the GW spectrum
is 
\bea
\label{eq_fiogw2}
\mathcal{{\bar{P}}}_h(\tau_0, k)
&=&
\left(
\frac{{\pi}A_B^2}{{128}
\,k_\star^2\,H_0^2\,\Omega_R}
\right)\,\left(1+\frac{k^2}{4\,k_\star^2}\right)^2\,
\left[
\mathrm{Ci}^2(| k \tau_R|)
+
\Big(\tfrac{\pi}{2}-\mathrm{Si}(| k \tau_R|)\Big)^2
\right]\,\Theta(2 k_\star-k)\,.
\nonumber\\
\eea
 This is an increasing function of $k/k_\star$,  dropping to
 zero for $k/k_\star\,>\,2$. As expected from
 the structure of the convolution integrals,
 its amplitude is proportional
 to the square of the amplitude
 of the magnetic
 field spectrum. Compare with Eq.~\eqref{cond_delsp}. It is also convenient to compute the energy density in GW, given by  
\begin{equation}
\label{Omega}
{\Omega_{\rm GW}\equiv\frac{1}{12}\left( \frac{k}{a H} \right)^2\bar{\mathcal{P}}_h}\,,
\end{equation}
being this quantity usually considered in GW experiments
aiming at characterising the SGWB spectral shape. It is expressed
in terms of GW frequency, $f/f_\star=k/k_{\star}$.

{Plugging  expression \eqref{eq_fiogw2} into \eqref{Omega} we get the form of the full GWs energy spectrum
\begin{align}
    \Omega_{\rm GW}(\tau_0, k)=
&\left(
\frac{\pi\,A_B^2}{128\,k_\star^2\,H_0^2\,\Omega_R}
\right)\left(\frac{k^2}{12H_0^2}\right)\left(1+\frac{k^2}{4\,k_\star^2}\right)^2
\left[
\mathrm{Ci}^2(| k \tau_R|)
+
\Big(\frac{\pi}{2}-\mathrm{Si}(| k \tau_R|)\Big)^2
\right] \nonumber \\
&\times \Theta(2 k_\star-k)\,,
\label{Omega_deltafunction}
\end{align}
where we used $a(\tau_0)=1$.

The plot of \eqref{Omega_deltafunction} as a function of $k/k_\star=f/f_\star$ for different values of $k_\star \tau_R$ is shown in figure \ref{delta}; we plot $\Omega_{\rm GW}(f)$ since this is the usually considered quantity in GW experiments aiming at characterising the SGWB spectral shape.   
Since a $\delta$-function source is physically hard to achieve,  we only focus on the qualitative shape of the spectrum. Therefore, we set the overall normalization parameters to unity, as they only change the amplitude of the function but not the dependence of $\Omega_{\rm GW}$ on $k/k_\star$; similarly, we fix $A_B=10^{-6}$. In our setup we assume $k_\star|\tau_R|\ll 1$, accordingly with the beyond--slow--roll magnetogenesis framework of \cite{Atkins:2025pvg}, where $k_\star\equiv-1/\tau_1$, $\tau_1$ is the instant at which the non--slow--roll phase begins during inflation, and $\tau_R$ denotes the end of inflation. With this choice, the only relevant parameter controlling the time integral part is the dimensionless combination $k_\star|\tau_R|$, which enters the argument of the oscillatory term through $k\tau_R=(k/k_\star)(k_\star\tau_R)$. For $k_\star|\tau_R|\ll1$ these oscillatory functions show a  mild variation. As a result, in the monochromatic delta-function case the spectrum is essentially controlled by the monotonically increasing prefactor and exhibits a sharp cutoff at $k=2k_\star$. This qualitative behaviour is similar to the monochromatic delta-function template discussed in \cite{Saga:2018ont} in an analogous setup of GWs induced by primordial magnetic fields, where a monotonically increasing $\Omega_{\rm GW}$ with a sharp cutoff at $k=2k_\star$ is likewise found.}

\begin{figure}[t!]
        \includegraphics[width=0.45\linewidth]{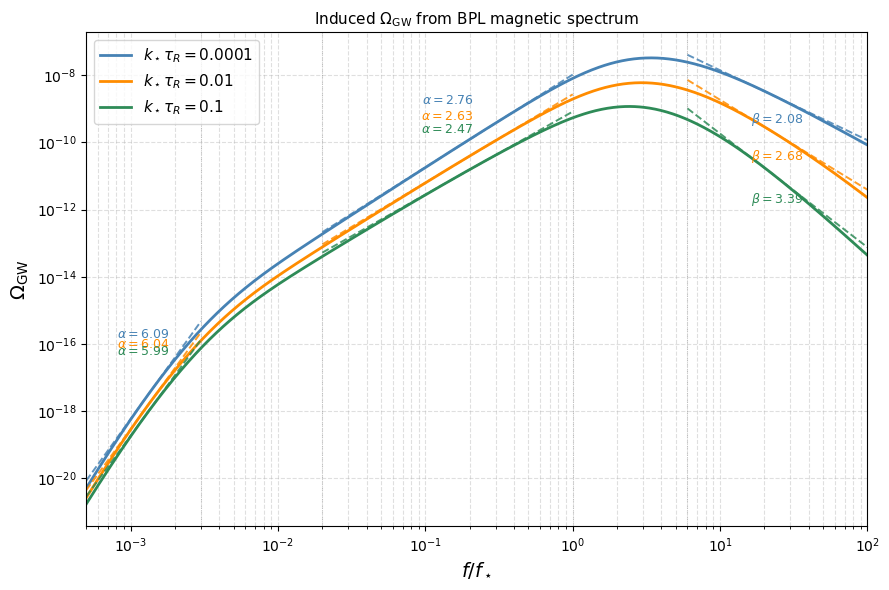}
        \,\,\,\,\,\,\,\,\,\,\,\,\,\,\, \includegraphics[width=0.45\linewidth]{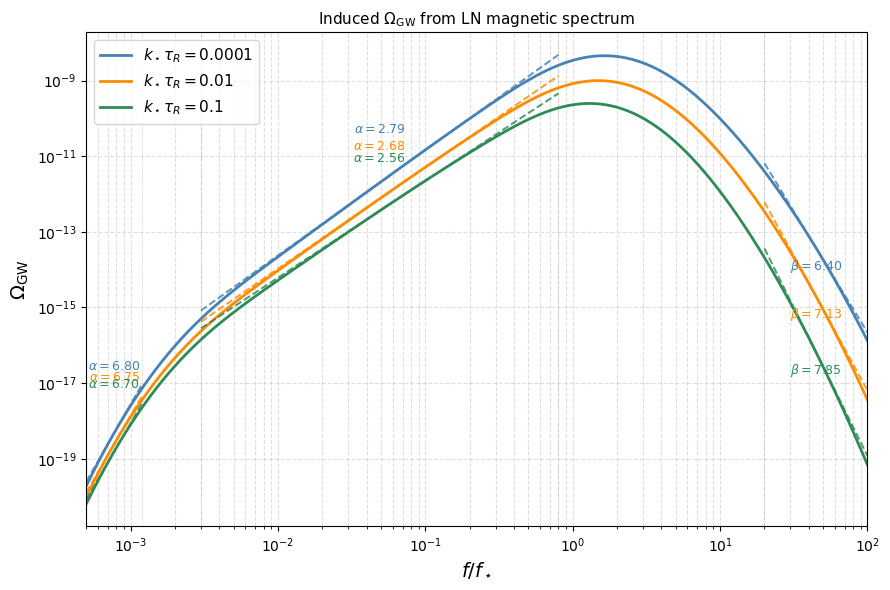}
    \caption{\small
    Profiles of $\Omega_{\rm GW}$ as function of GW frequency
    over a characteristic scale $f_\star$. {\bf Left panel}:
    GW spectrum induced by a broken power law profile
    for the magnetic field, \eqref{eq_bplp}, choosing  $n_1=4$, $n_2=-2$, $\sigma=1$. {\bf Right panel}: GW spectrum
    associated to the
    log-normal profile of Eq.~\eqref{eq_logp}, with $\Delta=0.8$. In both cases, $P_0=10^2$, and 
    fix the quantity $|k_\star \tau_R|=10^{-2}$ in Eq.\eqref{exp_ph3}.} 
    \label{fig_shap2}
\end{figure}

\medskip

After the previous analytical  model, we can
numerically handle  two more realistic templates
for the momentum profile of ${\cal P}_B$, which
are frequently considered in the analog context
of scalar induced gravitational waves. 
First, a broken power-law (BPL) profile
\begin{equation}
\label{eq_bplp}
{\cal P}_B(k)\,=\,P_0 \left( \frac{k}{k_\star}\right)^{n_1} \left(\frac12+\frac12 \left( \frac{k}{k_\star}\right)^\sigma \right)^{\frac{n_2-n_1}{\sigma}}
\end{equation}
where $P_0 $ an overall amplitude, $k_\star$ is a pivot momentum scale, $n_{1,2}$ characterize the slopes of the power-laws
on both sides of the break, and $\sigma$ controls
the smoothness of the break transition. Second,
a log-normal (LN) spectrum 
\begin{equation}
\label{eq_logp}
{\cal P}_B(k)\,=\,\frac{P_0}{\sqrt{2 \pi} \Delta}\,e^{-(\ln k/k_\star)^2/(2 \Delta^2)}
\end{equation}
maximised  at $k_\star$,
where $\Delta$ controls the width of the peak. We
perform  numerically the momentum
integrals ${\cal I}_B$, and represent in
Fig \ref{fig_shap2} our results
in terms of GW energy density of Eq.~\ref{Omega}.
 Both the panels in Fig \ref{fig_shap2} have a similar
profile, and we notice a rapid decay in power at $f>2 f_\star$,
analogously to what found for the case of delta-like magnetic
source. 
In both cases the GW spectrum exhibits the same qualitative structure:
an infrared rise at $k/k_\star\ll 1$, a maximum at wavenumbers of
order the characteristic scale $k_\star$, and a decay at larger
wavenumbers. At small $k$ the spectrum grows approximately as a
power law,
\begin{equation}
\Omega_{\rm GW}(k)\propto k^{\alpha(k)}\,,
\end{equation}
with an effective slope $\alpha(k)$ that is not constant and
progressively flattens as $k/k_\star$ approaches $10^{-1}$. Around
the peak the slope crosses zero, while at larger $k$ the spectrum
decreases and can be described by a negative effective slope,
$\Omega_{\rm GW}(k)\propto k^{-\beta(k)}$.
Figure~\ref{fig_shap2} shows the resulting $\Omega_{\rm GW}$ spectra
for the two magnetic templates considered (BPL and LN), together with
power-law fits performed over different frequency intervals. This
procedure highlights how the effective slopes $\alpha$ and $\beta$
vary across the integration ranges.
The main difference between the two cases appears in the ultraviolet
tail. For the BPL template the maximum occurs at $k/k_\star$ of order
a few and the subsequent decay is relatively mild, leaving an
extended high-$k$ tail consistent with a power-law behaviour. In
contrast, for the LN template the peak lies closer to
$k/k_\star\sim\mathcal{O}(1)$ and the ultraviolet suppression is
significantly steeper, reflecting the stronger localization of the
source spectrum around $k_\star$.
Finally, varying $k_\star\tau_R$ mainly rescales the overall
amplitude of the signal. In the range shown in the figure, increasing
$|k_\star\tau_R|$ reduces the amplitude of $\Omega_{\rm GW}$, in
agreement with the explicit dependence of the oscillatory functions
on $|k\tau_R|$, as already observed in the $\delta$-function case.
Finite-width peaked source spectra are more realistic than the
monochromatic approximation and naturally produce a GW signal
peaking around $k\sim k_\star$. Moreover, they display infrared and
ultraviolet behaviours that are qualitatively consistent with those
commonly found in peaked-source SGWB scenarios in the literature,
for instance in the case of scalar-induced gravitational waves (see
e.g.~\cite{Domenech:2021ztg} and references therein).

\section{
Computation of induced GW non-Gaussianity}
\label{sec_trispectrum}

\paragraph{Definitions and motivations}
 Since in our set-up the SGWB is sourced non-linearly 
 in fluctuations by primordial vector fields,  we
 expect that the corresponding connected higher order GW correlators are 
non-vanishing, and that they can lead to observational
consequences for GW experiments.  The physics
of dark sector involving vector fields, as discussed above, offers distinctive 
connections between magnetogenesis \cite{Durrer:1999bk,Caprini:2001nb,Mack:2001gc,Shaw:2009nf,Saga:2018ont,Ozsoy:2023gnl,Bhaumik:2025kuj,Maiti:2025cbi,Atkins:2025pvg}, dark matter \cite{Marriott-Best:2025sez}, 
and non-Gaussianities in the stochastic GW background. We aim to 
characterize the properties of the SGWB non-Gaussianities, with 
special emphasis on their implications for GW experiments which
we explore in the next Section.

In
this Section we demonstrate
that
the calculations carried on in Section~\ref{sec_setup} can be straightforwardly extended to study non-Gaussian GW
correlators. 
Since  three-point correlators of spin-2 GW
associated with   isotropic SGWB 
can {\it not} be directly detected with interferometer and pulsar timing array experiments (see e.g.  \cite{JimenezCruz:2025wqa})  we
focus on four-point correlators, and study
 the so-called GW trispectrum.
We show that the resulting  trispectrum in our
setup is stationary \cite{Powell:2019kid,Tasinato:2022xyq}
 and assumes a folded shape in terms of its momenta.

\begin{figure}[t!]
        \includegraphics[width=0.4\linewidth]{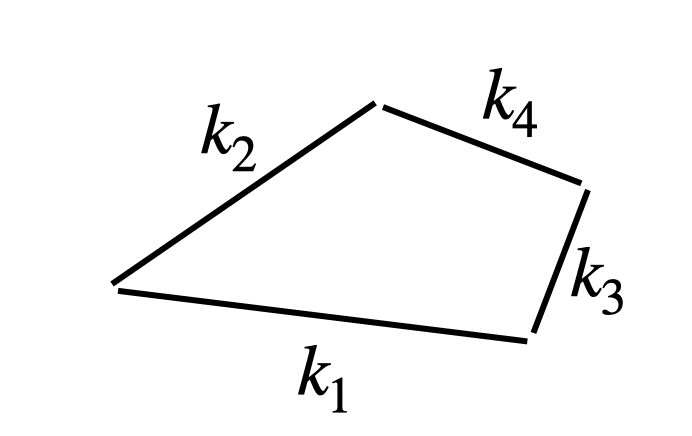}
        \,\,\,\,\,\,\,\,\,\,\,\,\,\,\, \includegraphics[width=0.4\linewidth]{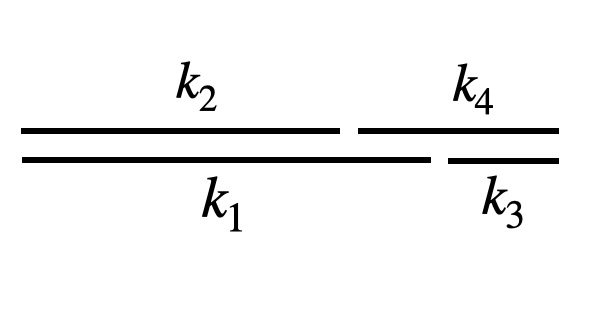}
    \caption{\small Graphical representation
    of the shapes of closed quadrilatera
    associated with the GW trispectrum. {\bf Left panel:} a generic closed shape. {\bf
    Right panel:} a flattened trispectrum shape, as the
    one we reduce to in our set-up. See main text for details.}
    \label{fig_quad}
\end{figure}

We will be interested on the connected
part of the GW four-point function, extracted
from the combination of correlators in momentum space
\be
\label{def_cT2}
\sum_{\lambda_i}\langle h^{(\lambda_1)}_{\bf k_1}(\tau)  \left(h^{(\lambda_2)}_{\bf k_2}(\tau) \right)^*
h^{(\lambda_3)}_{\bf k_3}(\tau)  \left(h^{(\lambda_4)}_{\bf k_4}(\tau) \right)^*
 \rangle_\Delta\,,
\ee
where $\lambda_{i}=\pm2$. In writing the previous equation 
we impose the momentum-conserving relation
\be
\label{cond_clquad}
\mathbf{k}_1+\mathbf{k}_3=\mathbf{k}_2+\mathbf{k}_4 \, 
\ee
associated with the rotational invariance of the background -- 
hence the momenta ${\bf k}_i$ form a closed quadrilaterum: see Fig~\ref{fig_quad}, left
panel. Moreover, 
as we are going to learn, in the set-up we consider the momenta ${\bf k}_i$ 
actually lie all {\it aligned} along a common direction ${\bf k}_i\,=\,k_i\,\hat n$. 
 See Fig \ref{fig_quad}, right panel for an example 
 of momentum alignment.

Hence the set-up we consider leads to shapes corresponding to flattened
quadrilateral 
configurations. 
Being the momenta of the massless spin-2 fields aligned,
 we impose the  conservation of total helicity in the correlators of Eq.~\eqref{def_cT2}. In Eq.~\eqref{def_cT2} 
 we then impose
{
 \be
 \lambda_1-\lambda_2+\lambda_3-\lambda_4=0
  \label{cond_pol}
 \ee}
 where $\lambda_i \,= \,
 \pm 2\,=\,(L,R)$.
Hence the $\lambda_i$ can assume the values 
{$(LLLL)$, $(LLRR)$, $(LRRL)$},
up to exchange of $L$ with $R$. In this
work, for definiteness, we assume that 
only the {$(LLLL)$} and {$(LLRR)$}  are turned on  (as
well as the corresponding contributions with $L\leftrightarrow R$)
and a sum over helicities reduce to
\begin{equation}
\sum_{\lambda_i}=\sum_{\lambda_{1,2}=\pm2}\,.
  \label{cond_pol2}
\end{equation}
We then define the trispectrum as
\be
\label{def_cT}
{\cal T}_h\,=\, {\frac{1}{4}}\frac{k_1^3 k_2^3 k_3^3}{({2} \pi^2)^3}\,
\sum_{\lambda_{1,2}=\pm2}\langle h^{(\lambda_1)}_{\bf k_1}(\tau)  \left(h^{(\lambda_1)}_{\bf k_2}(\tau) \right)^*
h^{(\lambda_2)}_{\bf k_3}(\tau)  \left(h^{(\lambda_2)}_{\bf k_4}(\tau) \right)^*
 \rangle_\Delta\,,
\ee
where $k_i^3$ overall factors are included to make
the result dimensionless, in analogy with the power
spectrum definition of Eq.~\eqref{def_tenps}. The structure of Eq.~\eqref{def_cT}
automatically satifies
the condition \eqref{cond_pol}.

\smallskip
From now on, we will be interested on the 
connected part of the trispectrum, which we still indicate with ${\cal T}_h$.
 For computing this quantity following  Eq.~\eqref{def_cT}, 
we start from the formal solution
of the GW mode function given in Eq.~\eqref{eq_frh}.
We proceed as in the previous section. The trispectrum ${\cal T}_h$ defined in Eq.\eqref{def_cT} results 
 factorizable  in a time and momentum integral
 as in Section \ref{sec_setup}:
\be
\label{eq_fatri2}
{\cal T}_h\,=\, \frac{16}{a^4(\tau)}\,J_\tau(\tau)\,J_B( k_i)
\ee
with 
\bea
\label{def_tintA}
J_\tau(\tau)\,=\,\prod_{i=1}^4\,\int d \tau_i\,\frac{
g_{k_i} (\tau,\tau_i)}{a(\tau_i)}
\eea
and
\begin{equation}
\label{def_tintB}
{J_B(k_i)\,=\,{\frac{1}{4}}
\frac{k_1^3 k_2^3 k_3^3}{({2} \pi^2)^3}\,\mathcal{C}^{(B)}(\mathbf{k}_i)}\,,
\end{equation}
with a polarization dependence
as explained after Eq.~\eqref{cond_pol} and where
\begin{eqnarray}
  &&\mathcal{C}^{(B)}(\mathbf{k}_i)\,=\,
\nonumber
\\
&&
\hskip-0.5cm
\sum_{\lambda_{1,2}=\pm2}\langle e^{(\lambda_1)\,ab}(\hat{{ k}_1})\,
    \tau^{(B)}_{ab}(\mathbf{k}_1)
    e^{(-\lambda_1)\,cd}(\hat{{k}_2})\,
    \tau^{(B)\,*}_{cd}(\mathbf{k}_2)
    e^{(\lambda_2)\,ef}(\hat{{k}_3})\,
    \tau^{(B)}_{ef}(\mathbf{k}_3)
    e^{(-\lambda_2)\,gh}(\hat{{k}_4})\,
    \tau^{(B)\,*}_{gh}(\mathbf{k}_4)\rangle_\Delta
\nonumber
\\
    \label{trisp}
\end{eqnarray}
We start with evaluating the time integral \eqref{def_tintA}, which -- 
as we are going to learn --  is responsible for forcing the aforementioned condition of alignment on the directions ${\bf k}_i$.

\paragraph{The time integral in Eq.~\eqref{def_tintA}.}

We evaluate Eq.~\eqref{def_tintA} 
 in the limit of large observation time $\tau$.
 Throughout, the Fourier momenta satisfy the closed-quadrilateral condition \eqref{cond_clquad}. 
The retarded Green function for each tensor mode is
\begin{equation}
g_{k_j}(\tau,\tau_j)
=
\frac{i}{2k_j}
\left(
e^{-ik_j(\tau-\tau_j)}-e^{ik_j(\tau-\tau_j)}
\right) .
\label{eq:green_time}
\end{equation}
Substituting Eq.~\eqref{eq:green_time} into Eq.~\eqref{def_tintA} yields a sum of $2^4=16$ oscillatory contributions, each proportional to exponentials of the form
$\prod_{j=1}^{4}\exp\!\left[\pm i k_j {\tau}\right]$.

In the late-time limit $\tau\to\infty$, all terms are rapidly oscillating and hence suppressed --  except for those for which the total phase is stationary. (Stationary phase approximation.)  
A direct inspection shows that stationarity is achieved  if the \emph{magnitudes} of the external momenta --  and not only combinations with their directions as Eq.~\eqref{cond_clquad} --  satisfy a relation of the form
\begin{equation}
\sum_{j=1}^{4} {s_j} k_j = 0 ,
\qquad
{s_j=\pm1} .
\label{eq:stationary_condition}
\end{equation}
Thus, configurations obeying the vectorial closure condition \eqref{cond_clquad} contribute at late times only if they correspond to \emph{flattened} or \emph{folded} trispectrum shapes, in which the four sides are superimposed.\footnote{Another possibility arises when the quadrilateral has two pairs of parallel sides; we defer a discussion of this case to future work.} See Fig.~\ref{fig_quad}, right panel,
as well as Appendix \ref{app_time} for technical details.

In other words, in our setup the surviving contributions correspond to configurations in which the four momenta are collinear and aligned along a common direction $\hat{\mathbf{n}}$. 
Each momentum can then be parametrized as
\begin{equation}
\mathbf{k}_i = s_i\, k_i\, \hat{\mathbf{n}},
\qquad
s_i=\pm1,
\qquad
k_i>0 ,
\label{eq_collinear_decomp}
\end{equation}
and the quadrilateral condition \eqref{cond_clquad} reduces to a scalar constraint as Eq.~\eqref{eq_collinear_decomp}:
\begin{equation}
s_1 k_1 + s_3 k_3 = s_2 k_2 + s_4 k_4 \, .
\label{eq:folded_scalar}
\end{equation}
See for example Fig.\ref{fig_quad}, right panel, for an example where the $s_i=1$
in the formula above.

\smallskip

We refer to this requirement as \emph{stationarity} of the trispectrum, and stress that it is an unavoidable consequence of the structure of the time integrals in our framework. The stationary condition
is essential for building observables
detectable with GW experiments. Once this condition is satisfied, the decorrelation arguments of Refs.~\cite{Bartolo:2018rku,Bartolo:2018evs} no longer apply, as discussed in~\cite{Powell:2019kid,Tasinato:2022xyq}. Indeed, the decoherence effects identified in~\cite{Bartolo:2018rku,Bartolo:2018evs} typically arise from phase misalignment among waves produced in multiple causally disconnected regions; through the central limit theorem, this mechanism drives the signal toward Gaussianity. In contrast, the alignment condition in our setup enforces that the GW  originate from the same direction, preventing such phase averaging. Our notion of \emph{stationary GW non-Gaussianity} therefore singles out a distinct class of tensor non-Gaussian signatures with support in these aligned configurations.

Independently of this effect,  \cite{Bartolo:2018rku,Bartolo:2018evs}  show that gravitational waves may accumulate random phases during their  propagation from source to detector, induced by long-wavelength energy-density fluctuations. These phases
-- physically associated with Shapiro time-delays -- act on short-wavelength GW modes traveling over cosmological distances. As a result, they tend to suppress phase correlations in initially non-Gaussian fields, and they reduce the amplitude of connected tensor $n$-point functions for $n \ge 3$. Interestingly, in our case  the unequal-time  correlators depend  on time differences only,
thanks to the stationarity condition above (see e.g. \cite{Powell:2019kid}).  Consequently, they are insensitive to the full propagation history of the waves and probe only the relatively short time scales relevant to the experiment.

Under
the condition~\eqref{eq_collinear_decomp}, the integral \eqref{def_tintA} can be evaluated
straightforwardly (see Appendix
\ref{app_time}), but the result is  cumbersome. For the simplest
choice $s_i=1$
 in Eq.~\eqref{eq:folded_scalar}, and for equal $k_i=k$ (a special case
 relevant for the examples we consider) we find

\be
\label{val_itau2}
\bar{{J}}_\tau
=
\frac{1}{8k^4 H_0^4\,\Omega^2_{R}}
\left[
\mathrm{Ci}^2(| k\tau_R |)
+
\Big(\tfrac{\pi}{2}-\mathrm{Si}(| k\tau_R |)\Big)^2
\right]^2.
\ee
It is simply proportional to the square
of the result we found for the two-point
function case: compare with  Eq.~\eqref{val_itau}.
{Having introduced the averaged time integral $\bar{J}_{\tau}$, we can accordingly define the averaged trispectrum, analogously as what we have done for the two-point function case
\begin{equation}
    {\cal \bar{T}}_h(k_i)\,=\, \frac{1}{4}\frac{16}{a^4(\tau)}\,\bar{J}_\tau\,J_B( k_i)\,.
    \label{trisp_ave}
\end{equation}}

\paragraph{The momentum integral of Eq.~\eqref{def_tintB}.}

We can now impose the collinearity condition of Eq.~\eqref{eq_collinear_decomp}, and express the
quantity $J_B({ k}_i)$ of Eq.~\eqref{def_tintB}
as (recall our choice of polarization
indexes in the definition of trispectrum \eqref{def_cT})
\begin{align}
J_B({ k}_i)
&= {\frac{1}{4(4\pi)^4}}
\left( \frac{k_1^3 k_2^3 k_3^3}{({2} \pi^2)^3} \right)
K_{i_1 j_1 \dots i_4 j_4}(\mathbf{k}_i)\,
\sum_{\lambda_{1,2}=\pm 2}
e^{(\lambda_1)}_{i_1 j_1}(\hat{{n}})
\, e^{(-\lambda_1)}_{i_2 j_2}(\hat{{n}}) \,
e^{(\lambda_2)}_{i_3 j_3}(\hat{{n}})
\, e^{(-\lambda_2)}_{i_4 j_4}(\hat{{n}})  \, ,
\nonumber
\\
&= {\frac{1}{4(4\pi)^4}}
\left( \frac{k_1^3 k_2^3 k_3^3}{(2 \pi^2)^3} \right)
K_{i_1 j_1 \dots i_4 j_4}(\mathbf{k}_i)\,
\Lambda_{i_1 j_1 i_2 j_2}
\Lambda_{i_3 j_3 ij_4}\,,
\label{eq:I4_kernel}
\end{align}
where we used the properties of combinations
of polarization tensors -- see Eq.~\eqref{def_ofLA}. 
The eight-index tensor in Eq.~\eqref{eq:I4_kernel}, after defining ${\bf q}_a={k}_a {\bf \hat n}-\mathbf{p}_a$,
results
\begin{align}
K_{i_1 j_1 \dots i_4 j_4}(\mathbf{k}_i)
&=
\int \prod_{a=1}^4 \frac{d^3 p_a}{(2\pi)^3}
\nonumber\\
&\hspace{0.3cm}
\Big\langle
B_{i_1}(\mathbf{p}_1)\,B_{j_1}( {\bf q}_1)
B_{i_2}^*(\mathbf{p}_2)\,B_{j_2}^*( {\bf q}_2)
B_{i_3}(\mathbf{p}_3)\,B_{j_3}({\bf q}_3)
B_{i_4}^*(\mathbf{p}_4)\,B_{j_4}^*({\bf q}_4)
\Big\rangle .
\label{eq:kernel_8pt}
\end{align}

\begin{figure}[t!]
    \centering
        \includegraphics[width=0.3\linewidth]{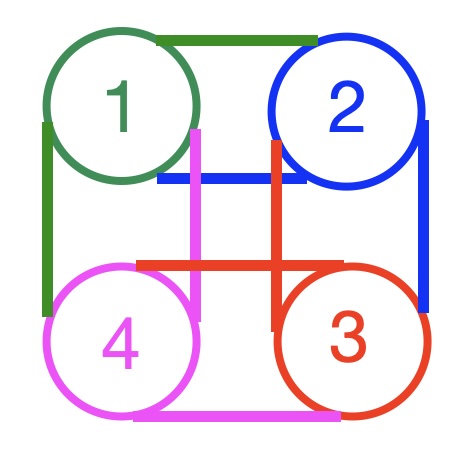}
    \caption{\small { A particular lay-out for the four building blocks
    of magnetic fields leading to  connected GW four-point functions, located
    on the vertexes of a square
    and linked by vector-field correlations. See main text for details.}}
    \label{fig_box}
\end{figure}

{ From this expansion, we are only interested in extracting the connected contributions, that correspond to the fully linked contractions in which all four source functions 
$S_1, S_2^*, S_3, S_4^*$ are connected through magnetic field pairings. On the contrary, the disconnected contributions correspond to terms that factorize into products of lower-order point functions, like
$\langle S_1 S_2^* \rangle \langle S_3 S_4^* \rangle$ or $\langle S_1  \rangle \langle S_2^*S_3 S_4^* \rangle$.

In order to isolate the connected part, we proceed as follows: we group the eight magnetic fields appearing in eq. \eqref{eq:kernel_8pt} into four quadratic blocks,
each block containing two magnetic Fourier modes, that we call legs.
At vertex $a$ we have the two legs $B_{i_a}(\mathbf{p}_a)$ and $B_{j_a}(\mathbf{q}_a)$, and analogously for the complex-conjugated blocks appearing in the second and fourth sources.
A Wick contraction pairs any legs, irrespective of whether their indices originate from an
$i$-slot or a $j$-slot of a given block.  For instance, $B_{i_1}(\mathbf{p}_1)$ can contract with
$B^*_{i_2}(\mathbf{p}_2)$ just as well as with $B_{j_2}^*(\mathbf{q}_2)$, and in the same way also two unconjugated modes can be connected, like $B_{i_1}(\mathbf{p}_1)$ and $B_{i_3}(\mathbf{p}_3)$. Recall that we can always pass from the conjugated to the non conjugated field by the relation $B_{i}(\mathbf{p})^*=B_{i}(-\mathbf{p})$.
We can thus define the contraction between two modes as
\bea
\Pi_{lm}&=&B_l(\mathbf{k}) B^*_m(\mathbf{k}')
\hskip0.3cm\Rightarrow\hskip0.3cm
\langle \Pi_{lm} \rangle\,=\,
 (2\pi)^3 \delta^{(3)}(\mathbf{k}-\mathbf{k}') P_B(k) \,\pi_{lm}({\hat k})
\,,
 \nonumber
\eea
and analogously between two non-complex conjugated modes, with a $+$ inside the $\delta$ function.
We now place the four source blocks (labelled by $a=1,\dots,4$) on the vertices of a square, and
represent each Wick contraction of magnetic modes as a line connecting two vertices (or better, two modes within two different vertices).
Disconnected contributions correspond to configurations in which the graph factorizes into two
(or more) independent subgraphs; these terms reduce to products of lower-order correlators and hence do
not contribute to the connected four-point function.
The connected contribution is obtained by retaining only those patterns that link all four vertices
into a single closed loop.
Therefore, there are six independent possibilities of connecting the blocks:
\begin{equation*}
\langle{\text{8pt function}}\rangle_{\rm conn}:\qquad
(1234)+ (1243)+ (1324)+ (1342)+ (1423)+ (1432)\,,
\end{equation*}
where the four numbers are the vertex labels $a$.
For any of these six contributions, a specific contraction is obtained by assigning, at each vertex,
which of its two legs is contracted to the next vertex along the cycle: for each of the six possibilities above, there are eight ways to contract magnetic Fourier modes among blocks. 

For example, focusing on the option indicated
as $(1234)$, one of the eight possible diagrams  
(the other choices lead to similar structures) is
\bea
(1234)^{1} &=& \int {\prod_{a=1}^4}
\,\frac{d^3 p_a}{(2\pi)^3}
\langle B_{i_1}({\bf p}_1)  B_{j_4}^\dagger({\bf q}_4) \rangle
\langle B_{j_1}({\bf q}_1)  B_{i_2}^\dagger({\bf p}_2) \rangle
\langle B_{j_2}^\dagger({\bf q}_2)  B_{i_3}({\bf p}_3) \rangle
\langle B_{j_3}({\bf q}_3) B_{i_4}^\dagger({\bf p}_4) \rangle
\nonumber
\\
&=&\,\delta[{\hat n} \left({ k}_3-{ k}_2+{ k}_1-{ k}_4 \right)]
\,\int d^3 p_1
P_B(p_1)\,
P_B(k_1-p_1)
P_B(p_1-k_1+k_2) 
P_B(k_4-p_1)
\nonumber\\
&&
\times \pi_{i_1 j_4}(p_1)
\pi_{i_2 j_1}(k_1-p_1)
\pi_{i_3 j_2}(p_1-k_1+k_2 )
\pi_{i_4 j_3}(k_4-p_1)\,,
\eea
where in the second equality we integrate over momenta
${\bf p}_{2,3,4}$, and for simplicity we 
understand the hats in the arguments of the $\pi_{ij}$
tensors. For this specific contribution,
the integral to compute for obtaining the 
corresponding part of $J_B$ (see Eq.~\eqref{eq:I4_kernel}) results
\be
\label{int_jb}
J^{{1}}_B\,=\,\frac{\left(k_1 k_2 k_3\right)^3}{{256 \pi}}\int \frac{dp\,d \mu}{p}\,C_1(k_i,p)\,
{\cal P}_B(p)\,
\frac{{\cal P}_B(k_1-p)}{|k_1-p|^3}
\frac{{\cal P}_B(p-k_1+k_2) }{|p-k_1+k_2|^3}
\frac{ {\cal P}_B(k_4-p)}{|k_4-p|^3} 
\ee
with
\be
\mu \equiv \frac{\mathbf{k}_1\cdot\mathbf{p}}{k_1 p}.
\ee
and
\bea
C_1(k_i,p)&=&
\Lambda_{i_1  j_1 i_2 j_2}( n)
\Lambda_{i_3 j_3 i_4  j_4}( n)
\,\pi_{i_1 j_4}(p)
\pi_{i_2 j_1}(k_1-p)
\nonumber
\\
&&\times 
\pi_{i_3 j_2}(p-k_1+k_2 )
\pi_{i_4 j_3}(k_1+k_3-k_2-p)
\label{res_c1}
\eea
Notice that, since we integrate over all directions $\hat p$, the 
quantity $J_B$ (hence the total trispectrum), results independent
on the direction $\hat n$ along which the momenta are aligned.  
\paragraph{An explicit, simple example.}
So far, our results are  general.
As done in Section \ref{sec_setup}, for
carrying on the integrals analytically we assume a delta-like source for the magnetic
field spectrum 
\be
\label{cond_delsp2}
\mathcal{P}_B(k)=A_B\,\delta\left(\ln\left[k/k_\star\right] \right)\,,
\ee
We introduce convenient variables 
\begin{eqnarray}
v &\equiv& \frac{p}{k_1},\qquad 
u \equiv \frac{|k_1\,\hat n-\mathbf{p}|}{k_1}\,,\qquad 
\sigma \equiv \frac{k_\star}{k_1}\,,
\\
u_a&\equiv&\frac{|k_1\,\hat n-k_2\,\hat n-\mathbf{p}|}{k_1},\qquad
u_b \equiv \frac{|k_4\,\hat n-\mathbf{p}|}{k_1},
\end{eqnarray}
so $\mu=(1+v^2-u^2)/(2 v)$. Using the properties
of Dirac delta function we can re-express
the power spectra as
\be
\mathcal{P}_B(p)\,=\,A_B\,\delta\left(\ln\left[p/k_\star\right] \right)\,=\,A_B\,\delta\left(\ln\left[v/\sigma\right] \right)
\,=\,A_B\,{\sigma}\,\delta\left(v-\sigma \right)
\,,\ee
and analog formulas apply to the
remaining contributions to Eq.~\eqref{int_jb}.
Eq~\eqref{int_jb} can then be re-expressed
as
\be
\label{int_jb2}
J_B^{{1}}\,=\,\frac{A_B^4}{
{256\pi}} \left(\frac{ k_2 k_3}{k_1^2}\right)^3\int {dv\,d \mu}\,C_1(k_i,p)\,
\delta\left(v-\sigma \right)\,
 \frac{\delta\left(u-\sigma \right) }{u^2}
\frac{\delta\left(u_a-\sigma \right) }{u_a^2}
\frac{\delta\left(u_b-\sigma \right) }{u_b^2}
\ee
In order for the delta-functions not to overconstrain
the system, we 
select all sizes of the momenta to be equal, $k_i=k_1$ for $i=2, 3, 4$,
so that two of the delta-function
conditions become redundant. 
The remaining delta-functions impose then
$u=v=\sigma$, and $\mu=1/(2 \sigma)$. 
Under these conditions, the function
 ${\cal C}_1$
in Eq.~\eqref{res_c1} reduces to
\begin{eqnarray}
{\cal C}_1&=&\frac{(1+{\mu^4}) (2-4 v \mu+v^2 (1+\mu^2))^2}{4 (1+v^2 -2 v \mu)}
\nonumber
\\
&=&\frac14 
\left(1+\frac{1}{4 \,\sigma^2} \right)^2
\left(1+\frac{1}{16 \,\sigma^4} \right)\,.
\end{eqnarray}
The integral can then be easily evaluated
giving
\be
\label{int_jb3}
J_B^{{1}}\,=\,\frac{A_B^4\,k_1^6}{{1024 \,\pi}\, k_\star^6} 
\left(1+\frac{k_1^2}{4 \,k_\star^2} \right)^2
\left(1+\frac{k_1^4}{16 \,k_\star^4} \right)
\,\Theta(2 k_\star-k_1)\,.
\ee
We can now assemble the previous result
\eqref{int_jb3}
with the time integral given in Eq.~\eqref{val_itau2}
 (in our
 situation with all the $k_i$ identical,
 as required by the delta-function
 conditions). The resulting contribution
 to the total 
 trispectrum in radiation domination
 is given by Eq.~\eqref{eq_fatri2}:
 when evaluated today, we multiply 
 the result by the transfer function 
 coefficient $\left( a^2(\tau)/2\right)^2$. 
Hence, 
the result is a function 
of a single momentum scale $k_1$
\begin{eqnarray}
{\bar{\mathcal{T}}^1}_h
&=&
\frac{A_B^4}{{8192\, \pi}\,k_\star^4 H_0^4\,\Omega^2_{R}}
\left[
\mathrm{Ci}^2(| k_1 \tau_R |)
+
\Big(\tfrac{\pi}{2}-\mathrm{Si}(| k_1 \tau_R |)\Big)^2
\right]^2
\,\frac{k_1^2}{ k_\star^2} 
\left(1+\frac{k_1^2}{4 \,k_\star^2} \right)^2
\left(1+\frac{k_1^4}{16 \,k_\star^4} \right)
\nonumber
\\
&=&{\frac{2}{\pi^3}}\,{{\bar{\cal P}}}_h^2(k_1)
\,\frac{k_1^2}{ k_\star^2} 
\left(1+\frac{k_1^2}{4 \,k_\star^2} \right)^{-2}
\left(1+\frac{k_1^4}{16 \,k_\star^4} \right)\,.
\end{eqnarray}
under our hypothesis of sharply peaked magnetic
spectrum. 
The amplitude of the contribution we computed for the GW trispectrum in the flattened shape is then proportional to the square of the 
GW power spectrum, times a simple polynomial function of the momentum scale. The other
contributions to $J_B$ from the aforementioned
permutations, being associated to similar diagrams,  are expected to give similar results. 
 {The presence of the $\delta$-function magnetic spectrum leads to a dependence on $k/k_\star$ which is analogous to the one observed for the two-point function case, where the signal grows and then exhibits a sharp cutoff for $k/k_\star=2$. 
 Therefore, the present discussion, although   preliminary,  already illustrates the mechanism by which the magnetic source shapes the GW trispectrum and the associated momentum dependence in this particular case.}

\paragraph{Summary of this Section}
Let us summarize
the results of this theoretical Section. We find
that vector-induced GW backgrounds are characterized
by intrinsic stationary  non-Gaussianities. The connected
part of the primordial
trispectrum acquires an amplitude proportional
to the square of GW power spectrum, and a shape 
corresponding to a  flattened quadrilateral shape with 
all sides superimposed: see Fig.~\ref{fig_quad}, right panel.

The scaling of the connected trispectrum amplitude as ${\cal P}_h^2$—rather than ${\cal P}_h^3$, as would be expected for a standard ``local-type'' expansion of GW non-Gaussian mode functions—implies that its magnitude can be parametrically enhanced. The characteristic folded shape of the trispectrum motivates the notion of \emph{stationary GW non-Gaussianity}, introduced in Section~\ref{sec_trispectrum} as a mechanism to address the detectability challenges emphasized in~\cite{Bartolo:2018rku,Bartolo:2018evs}. 
In the specific case of a delta-function magnetic power spectrum source, the momentum dependence of the trispectrum can be derived analytically, yielding a flattened quadrilateral configuration with equal sides. For more general sources, we expect a broader range of momentum dependences, while still preserving the underlying folded quadrilateral geometry. We now turn to the observational implications of these results and analyze their phenomenological consequences for gravitational-wave experiments.

\section{
Consequences for GW experiments}
\label{sec_det}

\paragraph{Motivations}  
We now investigate  implications for GW experiments of  non-Gaussian correlators 
in the SGWB from the early universe, with the geometrical characteristics
analysed in the previous Sections -- i.e.  a flattened 
shape for the trispectrum, whose momenta are aligned
along a common direction $\hat n$ and their size  satisfies
\begin{equation}
\label{cond_mom2}
k_1+k_3=k_2+k_4
\end{equation}
The polarization indexes in circular basis are such that the four-point correlator preserves helicity, see Eq. \eqref{def_cT}.
There are two complementary strategies to pursue  to test GW non-Gaussianities:
\begin{itemize}
\item We may examine how a sizeable  GW 
trispectrum in a flattened configuration \emph{modulates} the statistical properties of the
two-point function and of the gravitational-wave power spectrum,
thereby inducing distinctive and potentially observable signatures. 
\item Alternatively, we can study the
{\it direct detectability} of higher-order correlators,
such as the GW four-point function, by analyzing the response
of GW  experiments to this observable, and assessing the
corresponding detection prospects. Since
the trispectrum amplitude scales as ${\cal T}_h \propto {\cal P}_h^2$, the detectability of
this quantity depends on the values of the GW
two-point correlator. 
\end{itemize}

We consider examples
of both approaches, respectively in Section~\ref{sec_pta} and \ref{sec_gbd}.
We
 begin by presenting a schematic and general discussion of the underlying
physical effects we are going to investigate. 

\smallskip

A typical
GW signal $s_A$ measured by a GW experiment 
 labeled with $A$ is given by the contraction of the GW $h_{ij}(\tau, {\bf x})$ 
with the so-called detector tensor $d_{A}^{ij}$
\be
\label{def_sig}
s_A\,=\,h_{ij} d_{A}^{ij}\,,
\ee
 the latter  quantity depends
on the geometric configuration of the GW detector,  and its properties.
We assume the corresponding measurement to be made at time $\tau$
and at position ${\bf x}_A$~\footnote{More realistically, as we will learn in the next subsections, the results
depend on various locations as the Earth and pulsar positions in pulsar timing arrays or the vertexes of interferometer
arms. We adopt this preliminary simplification to make our arguments more
transparent.}.
We define
the combination
\be
d_A^{(\lambda)}(\hat n)\,=\,e_{ij}^{(\lambda)}(\hat
n)\,d_{A}^{ij}\,,
\ee
with $\hat n$ the GW direction. 
Computing the signal two-point function, and passing
to Fourier space, we have
\bea
\langle s_A\left(\tau, {\bf x}_A\right) s_B\left(\tau, {\bf x}_B\right) \rangle
&=&\sum_\lambda\,\int k^2 dk \,\frac{d^2 \hat{n}}{(2 \pi)^3}\,
d_A^{(\lambda)}(\hat n) d_B^{(-\lambda)}(\hat n)\,e^{i k\,\hat n\,({\bf x}_A-{\bf x}_B)}\, \langle h_{\bf k}^{(\lambda)} h_{\bf k}^{(-\lambda)} 
\rangle_\Delta
\nonumber
\\
&=& {2}\int d \ln k\,\gamma_{AB}(k)\,{\cal P}_h(k)\,,
\eea
where 
\bea
\label{def_orf2}
\gamma_{AB}\,=\,\frac{1}{4 \pi} \int d^2 \hat{n}\,\,
\left(\sum_\lambda
d_A^{(\lambda)}(\hat n) d_B^{(-\lambda)}(\hat n) \right)\,e^{i k\,\hat n\,({\bf x}_A-{\bf x}_B)}
\eea
is the so-called {\it overlap reduction function} (ORF) which controls
how to signal two-point correlations depend on the detector properties
and configuration~\footnote{The overall factor 
controlling the ORF normalization in Eq.~\eqref{def_orf2} can change depending
on conventions for  each type of GW experiment. We will be explicit on
the normalization definitions in 
what comes next.}.

\smallskip
The ORF is the
bridge between theoretical predictions of GW properties --
as the GW power spectrum -- and experimental
quantities. Its features  allow us to investigate
whether the measured signal is due to a GW background. For
example,  data distributed following Hellings-Downs curve 
in pulsar timing array measurements (more on this in Section \ref{sec_pta}) offer circumstantial evidence for a GW
interpretation of the measured GW signal. 

\paragraph{The variance of the ORF.}
The ORF can  be interpreted
as arising from the {\it mean value} in an ensemble average of GW measurements. Hence
its properties  can be characterized by a {\it variance},
which should be taken into account when
confronting theory with experiment \cite{Allen:2022dzg}. There are several possible
contributions to the variance of an ORF, depending on the 
experiment under consideration.
Let us schematically
estimate the contribution of connected
four-point function to what was dubbed {\it total variance}~\footnote{Although
\cite{Bernardo:2022xzl} focuses on pulsar timing arrays, their definition
of total variance can be in principle applied to more general GW experiments.} of the ORF
in \cite{Bernardo:2022xzl}. It is defined as
\bea
\sigma_{AB}^2&=&\langle s_A  s_B s_A s_B \rangle
-\langle s_A  s_B \rangle^2\,.
\eea

Besides its disconnected parts, let us assume a non-vanishing connected contribution to the trispectrum, characterized by a flattened shape as discussed in Section \ref{sec_trispectrum}. This implies that all momenta
are forced to align along a common direction $\hat n$. 
The connected part of the signal four-point function -- associated with the connected trispectrum -- can then be expressed as
\bea
\label{eq_connsig4pt}
\langle s_A  s_B s_A s_B \rangle_{\rm{conn}}
\,=\,{\frac{1}{4 \pi^2}}\,\int {d \ln k_1}\, {d \ln k_2}\, {d \ln k_3} {d  k_4}\,
{\cal T}_h(k_i)\,\Sigma_{\rm{conn}}^2\,,
\eea
where (we use the specific trispectrum
definition of \eqref{def_cT})
\bea
\Sigma_{\rm{conn}}^2&=&\frac{1}{4\pi}
\int d^2 n d^2 n_2 d^2 n_3 d^2 n_4\,\delta(\hat n-\hat n_2)\,\delta(\hat n-\hat n_3)\,\delta(\hat n-\hat n_4)
\delta(k_1+k_3-k_2-k_4)
\nonumber
\\ 
&&\times\left(\sum_{\lambda_{1,2}=\pm2} 
d_A^{(\lambda_1)}(\hat n) d_B^{(-\lambda_1)}(\hat n)
d_A^{(\lambda_2)}(\hat n) d_B^{(-\lambda_2)}(\hat n)\right)
\,e^{i k_1 \hat n \cdot {\bf x}_A-i k_2 \hat n_2 \cdot {\bf x}_B+i k_3 \hat n_3 \cdot {\bf x}_A-i k_4 \hat n_4 \cdot {\bf x}_B}
\nonumber
\\ 
&=&\frac{1}{4 \pi} \int d^2 \hat{n}\,\,
\left(\sum_{\lambda_{1,2}=\pm2} 
d_A^{(\lambda_1)}(\hat n) d_B^{(-\lambda_1)}(\hat n)
d_A^{(\lambda_2)}(\hat n) d_B^{(-\lambda_2)}(\hat n)\right)\,e^{i (k_1+k_3)\,\hat n\,({\bf x}_A-{\bf x}_B)}\,,
\label{def_sigco}
\eea
and  we make use of the condition of alignment 
of all four momenta along a common direction $\hat n$,
and the sum over polarization indexes satisfy
the helicity conservation Eq~\eqref{cond_pol}.

Assembling the connected result in Eq.\eqref{eq_connsig4pt} with 
the disconnected contributions, we can easily compute the total
variance as
\bea
\sigma_{AB}^2&=&\langle s_A  s_B s_A s_B \rangle
-\langle s_A  s_B \rangle^2
\\
&=& \int d \ln k_1
d \ln k_2\, \,{\cal P}_h(k_1) 
{\cal P}_h(k_2)
 \left[
\Sigma_{\rm disc}^2(k_1, k_2) +
\int d \ln k_3 \frac{{\cal T}_h(k_1,k_2,k_3)}{{\cal P}_h(k_1) 
{\cal P}_h(k_2) } \,
\Sigma_{\rm conn}^2(k_1, k_2)
\right]
\nonumber
\\
\label{eq_varmod}
\eea
with
\bea
\Sigma^2_{\rm disc}(k_1,k_2)&=&
\gamma_{AB}(k_1)\gamma_{AB}(k_2)+ \gamma_{AA}(k_1) \gamma_{AA}(k_2)
\eea
and $\Sigma^2_{\rm conn}$ is given in Eq.~\eqref{def_sigco}. 
The first two terms within parenthesis 
of Eq.~\eqref{eq_varmod} depend on the disconnected
contributions to the four-point function,
while the last term depends on its connected
part. 
Hence, Eq.~\eqref{eq_varmod} indicates that
the connected GW trispectrum in a flattened configuration contributes
to the total variance of the two-point function.
Consequently, it
  potentially affects GW measurements.
  In the examples discussed in Section \ref{sec_trispectrum}, the
  amplitude
  of the trispectrum is proportional to the square of the 
  power spectrum amplitude, hence we can expect an order-one value for  
  the integral over $d \ln k_3$, contributing to the last term within 
  parenthesis of Eq.~\eqref{eq_varmod}. The contribution to the variance
  of the  connected trispectrum -- in such set-up -- should
  be comparable to the one of the disconnected
  parts of the four-point function.
  This fact 
is particularly relevant for pulsar timing arrays,
see Section \ref{sec_pta}.

\paragraph{The ORF for the four-point function of GW signals.}
 Besides a modulation of the GW two point function, 
 as a matter of principle a connected
 trispectrum can be measured
 directly by taking the four-point function of the GW signal, and extracting its connected component exploiting its associated ORF. Assuming that
 the trispectrum shape acquires only a flattened
 configuration, within the hypothesis discussed  above, we compute the
 associated four-point ORF through the formula 
 \begin{equation}
 \label{def_orffpf}
 \gamma_{ABCD}\,=\,
 \frac{1}{4 \pi} \int d^2 \hat{n}\,\,
\left(\sum_{\lambda_{1,2}=\pm2} 
d_A^{(\lambda_1)} d_B^{(-\lambda_1)}
d_A^{(\lambda_2)} d_B^{(-\lambda_2)}\right)\,e^{i k_1\,\hat n\,({\bf x}_A-{\bf x}_D)}
\,e^{-i k_2\,\hat n\,({\bf x}_B-{\bf x}_D)}\,e^{i k_3\,\hat n\,({\bf x}_C-{\bf x}_D)}
 \end{equation}
where all the $d's$ depend on the common GW direction $\hat n$ and  the $x$'s are the detector positions. 
 The polarization indexes satisfy the helicity
condition \eqref{cond_pol}, and the momenta -- aligned
along $\hat n$ -- satisfy the condition  \eqref{cond_mom2}. The previous angular integral leads to 
the four-point ORF and, as we shall see in Section~\ref{sec_gbd}, it can be performed
analytically -- at least in the case of ground-based detectors.

\subsection{
Pulsar timing arrays}
\label{sec_pta}

Pulsar timing arrays (PTA) offer a powerful
method for searching for the SGWB background
at nano-Hertz frequencies, which recently
lead to significant hints of detection \cite{NANOGrav:2023gor,Reardon:2023gzh,Xu:2023wog,EPTA:2023fyk}. 
Time delays on the pulsar periods, indicated
with $z(t)$, or the corresponding time residuals called $R(t)$, are used as indicators
of the presence of  a SGWB. We refer to \cite{Maggiore:2018sht} for a pedagogical discussion of basic
concepts and techniques in PTA physics. 
The recent preliminary measurements of 
the SGWB at nano-Hz frequencies might also be explained
by certain models of second order, induced GW (see e.g.
\cite{Figueroa:2023zhu}),
similar to the scenarios we consider in our work.
Hence it is very relevant to study implications of GW
non-Gaussianities in this context.

The  ORF corresponding to the PTA time-delay two-point
functions has been first computed by Hellings and Downs in \cite{Hellings:1983fr}.
 The detector tensor $d_A^{ij}$ of Eq.~\eqref{def_sig} reads in this contect
\begin{equation}
d_A^{ij}\,=\,\frac{n^i n^j}{2(1+\hat x_A\cdot \hat n)}
\end{equation}
where $\hat n$ is the GW direction, and $\hat x_A$ the direction of pulsar
$A$ with respect to the Earth position.  
The ORF  leading
to the Hellings-Downs curve correlate
GW signals measured with two pulsars $A$ and $B$. It 
is given by the integral in Eq.~\eqref{def_orf2}, sometimes multiplied
by a conventional normalization factor $3/4$ -- see e.g. \cite{Bernardo:2024bdc}. 
As customary, we neglect the so-called \emph{pulsar-term} contributions, which cancel upon integration unless two pulsars are coincident. Under this assumption, the resulting integral becomes independent of the specific pulsar locations, as well as on the GW frequency. It reads
\begin{equation}
\gamma_{AB}(\zeta) \;=\; \frac{3}{2}\,\frac{1-\cos\zeta}{2}\,\ln\!\left(\frac{1-\cos\zeta}{2}\right)
- \frac{1}{4}\,\frac{1-\cos\zeta}{2} + \frac{1}{2}\,\left(1+ \delta_{AB}\right).
\label{eq:HD_explicit}
\end{equation}
where  $\zeta$ is the angle between the vectors pointing from the Earth 
towards pulsar $A$ and $B$ respectively. In the limit
of coincident pulsars, $A=B$, the pulsar
term becomes important leading to the 
 Kronecker function $\delta_{AB}$ in Eq.\eqref{eq:HD_explicit}. See e.g. \cite{Romano:2023zhb}.

\smallskip

The Hellings-Downs curve is a central quantity in PTA measurements -- its
characteristics angular correlations is a crucial test for determining
the GW origin of PTA measurements. Recently, much work has been 
devoted in computing and physically characterizing the {\it variance} 
of 
Hellings-Downs correlators \cite{Allen:2022dzg,Bernardo:2022xzl,Allen:2022ksj,Bernardo:2023bqx,Romano:2023zhb,Valtolina:2023axp,Allen:2024rqk,Grimm:2024lfj,Agarwal:2024hlj,
Lamb:2024gbh,Wu:2024xkp,Pitrou:2024scp,Domcke:2025esw}. There are
several 
 different contributions
 to the variance of this curve. Following the arguments of Section~\ref{sec_trispectrum}, we are
  interested to the so-called total variance  (dubbed in this way in \cite{Bernardo:2022xzl}).
The total variance corresponds to the expected uncertainty for a {\it single pulsar pair} whose time delays are caused and correlated by the gravitational-wave background. It  is
associated with the four-point function of Eq.~\eqref{def_orffpf2}.
It
is constituted  by two pieces, a
disconnected part -- already discussed in the literature --
and a connected one -- which we include here for the first time.
The disconnected part is the total variance studied in \cite{Bernardo:2022xzl}, 
\be
\Sigma^2_{\rm disc}\,=\,\gamma_{AB}^2+\gamma_{AA}^2
\ee
which is easily computable from Eq.\eqref{eq:HD_explicit}. It coincides
with the results of \cite{Bernardo:2022xzl}.

\begin{figure}[t!]
    \centering
        \includegraphics[width=0.5\linewidth]{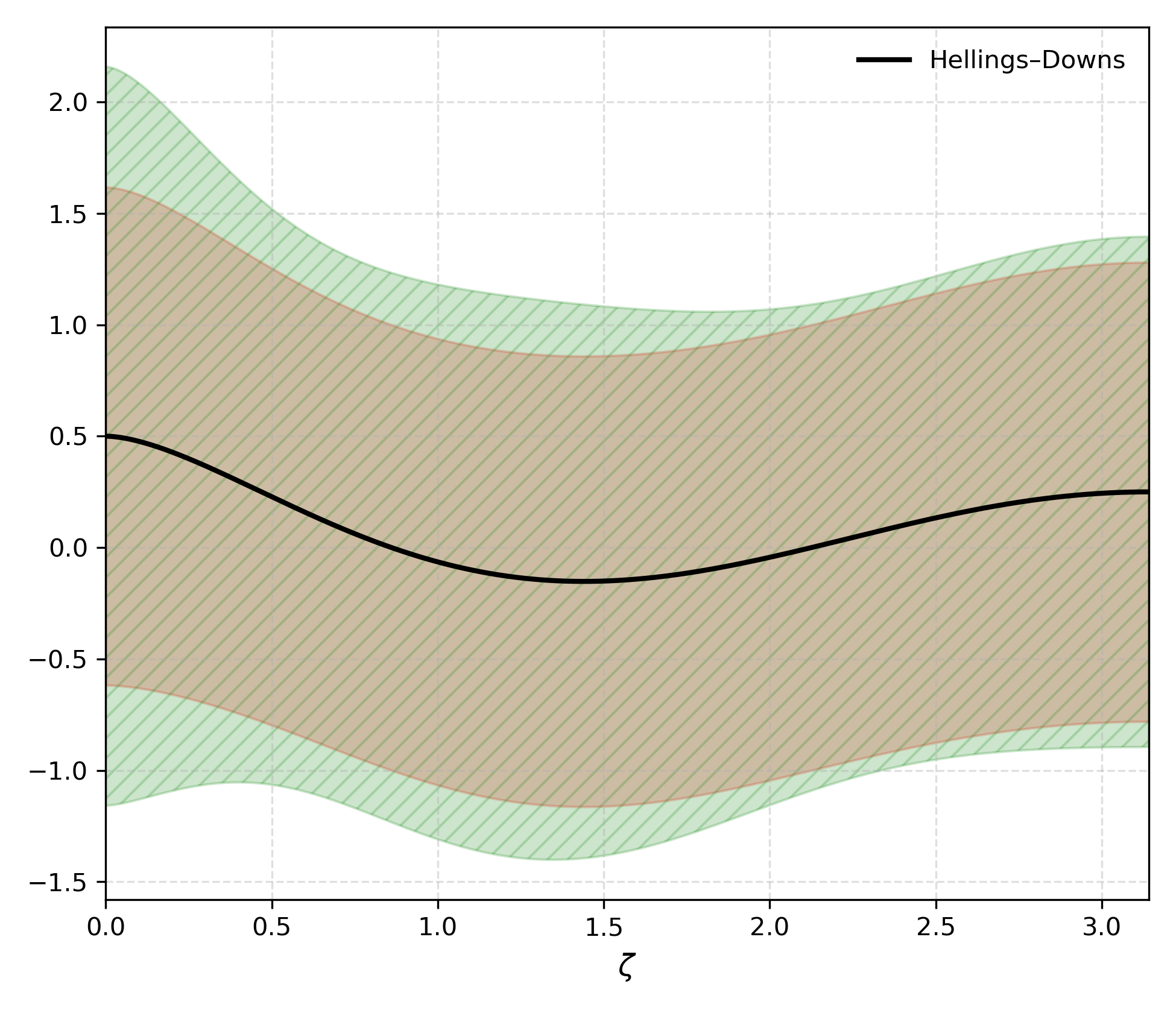}
    \caption{
    \small{Representation of the Hellings-Down
    function (black line); its  total variance associated
    with disconnected contributions to the signal
    four-point function (brown band);  finally,
    the total variance accounting for non-vanishing
    contributions of the connected GW trispectrum (green band). See main text for details.}}
    \label{fig_HDv2}
\end{figure}
We compute for the first time the connected part contribution --
last term of Eq.~\eqref{eq_varmod} and in particular the quantity in Eq~\eqref{def_sigco}, including an overall factor of $3/4$ to match with the
aforementioned Hellings-Downs normalization convention.
See also \cite{Tasinato:2022xyq} for another example of 
possible contribution
of GW four-point functions to Hellings-Downs--type
correlations.
Recall 
that we need to sum over all polarizations satisfying
Eq.~\eqref{cond_pol2},
and
we specialize to our definition
of trispectrum of Eq.\eqref{def_cT}.
 We find
\begin{eqnarray}
\Sigma^2_{\rm conn}&=&
\frac{3}{16 \pi}
\int d^2 n \left[
\left(\sum_{\lambda_{i}}
d_A^{(\lambda_1)}(\hat n) d_B^{(-\lambda_1)}(\hat n)
d_A^{(\lambda_2)}(\hat n) d_B^{(-\lambda_2)}(\hat n)\right)
\right]
\\
&=&\frac{1}{40} \left[6 + (134 - 139 y) y + 30 y (2 + 3 y) \ln{y}\right]\,,
\end{eqnarray}
where we call
$y=(1-\cos \zeta)/2$, and we denote $\left(d_A^{(L)} d_B^{(R)}\right)=d_A^{ij}
d_B^{pq}\,e_{ij}^{(L)}e_{pq}^{(R)}$. 
We plot in Fig.~\ref{fig_HDv2} the Hellings-Downs ORF $\gamma_{AB}(\zeta)$
as a function of the angle among pulsars. We include the effect of variances
in color bands. In red band, the disconnected contribution to the variance,  delimited within the region  $\gamma_{AB}\pm \Sigma_{\rm disc}$. (Compare with Fig 1 of \cite{Bernardo:2022xzl}). In green band, the new connected contribution, within the region $\gamma_{AB}\pm \sqrt{\Sigma_{\rm disc}^2+10\,\Sigma_{\rm conn}^2}$. (We set to 10  the integral over $\ln k_3$ in Eq. \eqref{eq_varmod}, in order to make more clear the angular dependence of the result.) 
We learn that the connected part of the GW non-Gaussianity
can increase the total variance of the Hellings-Downs
ORF, hence allowing for scattering of experimental
data around the mean Hellings-Downs ORF for single pulsar-pair measurements. It would be interesting
in the future to further study the observational implications
on PTA data
of the amplification effects on the size
of the variance  induced by the connected part of the non-Gaussian GW trispectrum.

\subsection{
Ground-based detectors}
\label{sec_gbd}

We now outline the analytical computation of the ORF four-point function defined in Eq.~\eqref{def_orffpf}, which is the key observable for directly probing the connected component of the signal four-point correlator. Improvements in the sensitivity of ground-based detector networks, both within the LVK array~\cite{LIGOScientific:2014pky,VIRGO:2014yos,Aso:2013eba} and in future facilities such as the Einstein Telescope~\cite{ET:2025xjr}, open new opportunities to detect primordial SGWBs in the frequency range of tens of Hertz, should such signals exist. Cosmological sources capable of producing backgrounds in this band, including second-order induced SGWBs, are reviewed in~\cite{Allen:1997ad}. It is therefore essential to investigate their detectability prospects in scenarios where the signal is non-Gaussian, as in the framework considered here. Although we do not present explicit detection forecasts—since these depend on the detailed frequency dependence of the four-point correlators—we emphasize that, because the trispectrum amplitude scales as ${\cal P}_h^2$, it can be significantly enhanced in frequency intervals where the gravitational-wave signal itself is amplified, as in the
second order GW induced scenarios explored in previous sections. 
 In our analysis of implications
 of GW non-Gaussianities, an advantage 
  of ground-based detectors is that the relevant integrals as Eq.\eqref{def_orffpf} can be evaluated analytically.
  For example, the ORF for correlations among two detectors
  is well-known, see e.g. \cite{Allen:1997ad}.
  Here we consider correlating measurements of a non-Gaussian stochastic gravitational-wave background (SGWB) across {\it four} distinct detectors—explicit examples to have in mind being LIGO Hanford, LIGO Livingston, Virgo, and KAGRA. In the presence of a connected gravitational-wave trispectrum, such a four-detector correlation is non-vanishing and, in principle, observable.

Assuming equal-length detector arms, the detector
tensor $d_A^{ij}$ reads in this case
\be
d_A^{ab}=\frac12 \left( u_A^a u_A^b -v_A^a v_A^b \right)
\ee
with ${\bf u}_A$, ${\bf v}_A$ denoting the directions in three-dimensional
space of the arms of the 
interferometer denoted with the index $A$. Notice the tracelessness condition
$d_A^{ii}=0$, as well as the fact that $d_A^{ij}$ is now 
independent from the GW direction $\hat n$ and momentum.
In this set-up, we can write Eq.~\eqref{def_orffpf} as follows
\begin{equation}
 \label{def_orffpf2}
 \gamma_{ABCD}(k_1,k_2)\,=\,
 \frac{d_A^{ab} d_B^{cd}  d_C^{ef} d_D^{gh}}{4 \pi} \int d^2 \hat{n}\,
 \Lambda_{abcd}(\hat n) \Lambda_{efgh}(\hat n)
\,e^{i k_1\,\hat n\,({\bf x}_A-{\bf x}_D)}
\,e^{-i k_2\,\hat n\,({\bf x}_B-{\bf x}_D)}
\,e^{i k_3\,\hat n\,({\bf x}_C-{\bf x}_D)}\,, \end{equation}
where we use identity \eqref{def_ofLA}, we impose the helicity  condition \eqref{cond_pol2}  and condition \eqref{cond_clquad} on GW momenta direction, as well as the 
  flattened shape condition
for the trispectum along a GW direction $\hat n$ (see Fig.~\ref{fig_quad}, right panel). 

\smallskip

To proceed, generalizing  \cite{Allen:1997ad} to the case
of connected four point functions, we introduce
the combinations
\begin{eqnarray}
\alpha&=&
k_1\,|{\bf x}_A-{\bf x}_D| - k_2\,|{\bf x}_B-{\bf x}_D|+ k_3\,|{\bf x}_C-{\bf x}_D|
\\
\hat s&=&\frac{ 
k_1\,({\bf x}_A-{\bf x}_D) - k_2\,({\bf x}_B-{\bf x}_D)+ k_3\,({\bf x}_C-{\bf x}_D)
}{ k_1\,|{\bf x}_A-{\bf x}_D| - k_2\,|{\bf x}_B-{\bf x}_D|+ k_3\,|{\bf x}_C-{\bf x}_D|}\,,
\end{eqnarray}
 which corresponds to a weighted size $\alpha$ of GW momenta, and an average direction $\hat s$.  We recall the relation Eq.~\eqref{def_ofLA},  between the projector tensor $\Lambda$ and the GW direction $\hat n$: once expanded over its contributions  it reads
\be
\begin{aligned}
   2 \Lambda_{abcd}=& \delta_{ad} \delta_{bc}+ \delta_{ac} \delta_{bd}- \delta_{ab} \delta_{cd} \\
    +&\delta_{cd} \hat{{n}}_a \hat{{n}}_b + \delta_{ab} \hat{{n}}_c \hat{{n}}_d- \delta_{bd} \hat{{n}}_a \hat{{n}}_c - \delta_{ac} \hat{{n}}_b \hat{{n}}_d -\delta_{bc} \hat{{n}}_a \hat{{n}}_d-\delta_{ad} \hat{{n}}_b \hat{{n}}_c \\
    +& \hat{{n}}_a \hat{{n}}_b \hat{{n}}_c \hat{{n}}_d. 
\end{aligned}
\label{eq:Lambda_exp}
\ee
It is then convenient to define the following combinations
controlling correlations among two detectors
\begin{eqnarray}
F^{(0)}_{AB}&=& \frac12 d_A^{ab}d_B^{cd}  \left(\delta_{ac} \delta_{bd}+{\rm perms} \right)\,,
\\
F^{(2)}_{AB}&=& -\frac12\,d_A^{ab}d_B^{cd}  \left(\delta_{ac} n_b n_d+{\rm perms} \right)\,,
\\
F^{(4)}_{AB}&=& \frac12\,d_A^{ab}d_B^{cd}  n_a n_b n_c n_d\,,
\end{eqnarray}
where the index $(p)$ indicates the power of 
unit vector
$\hat n$ involved. The integral to deal with
formally becomes
\begin{equation}
 \label{def_orffpf3}
 \gamma_{ABCD}\,=\,
 \frac{1}{4 \pi} \int d^2 \hat{n}\,
 \left(F_{AB}^{(0)}+F_{AB}^{(2)}+F_{AB}^{(4)} \right)
 \left(F_{CD}^{(0)}+F_{CD}^{(2)}+F_{CD}^{(4)} \right)
\,e^{i \alpha\,\hat s \cdot \hat n}\,.
 \end{equation}
We multiply the two parenthesis 
in Eq.~\eqref{def_orffpf3}, and define the following combinations 
\begin{eqnarray}
G^{(0)}_{ABCD}&=& F^{(0)}_{AB} F^{(0)}_{CD}
\,,
\\
G^{(2)}_{ABCD}&=& F^{(2)}_{AB} F^{(0)}_{CD}+F^{(0)}_{AB} F^{(2)}_{CD}
\,,
\\
\dots&=& \dots
\\
G^{(8)}_{ABCD}&=& F^{(4)}_{AB} F^{(4)}_{CD}\,,
\end{eqnarray}
which depend on all four detectors, and assemble
the powers of unit vector $\hat n$. We can now
 use the identity
\be
e^{i \alpha \hat n \cdot \hat s}\,=\,\sum_{\ell=0}^{\infty}
i^\ell\,(2 \ell+1)\,j_\ell (\alpha)\,P_\ell(\hat n \cdot \hat s)
\,,
\ee
with $j_\ell$ the spherical Bessel function, and $P_\ell$ 
the Legendre polynomial of order $\ell$. Plugging
into Eq.~\eqref{def_orffpf3} and expanding, we find
that ORF can be expressed as
\begin{eqnarray}
\gamma_{ABCD}\,=\,
 {\sum_{n=0}^4 I^{(2n)}}
\end{eqnarray}
where each $I^{(2n)}$ reads
\begin{eqnarray}
I^{(2n)}&=&\sum_{\ell=0}^n\,(-)^{\ell}\,
\frac{\left(4 \ell+1 \right)}{4 \pi}\,j_{2 \ell}(\alpha)
\int d^2 n\,P_{2\ell} (\hat n \cdot \hat s)\,G^{(2n)}(\hat n)\,.
\end{eqnarray}
Hence we reduce
the problem to compute integrals of Legendre polynomials, weighted by polynomials in $\hat n$ -- a straightforward operation
which can be carried on analytically. 

For example, let us work out explicitly the computation
for $I^{(0)}$
\begin{eqnarray}
I^{(0)}&=&\,\frac{j_{0}(\alpha)}{4\pi}
\int d^2 n\,P_{0} (\hat n \cdot \hat s)\,G^{(0)}(\hat n)
\\
&=&\frac{j_{0}(\alpha)}{4\pi} \left( d_A^{ab} d_B^{ab} \right)
 \left( d_C^{cd} d_D^{cd} \right)
 \int d^2 n\,P_{0} (\hat n \cdot \hat s)
 \\
 &=&\frac{\sin{\alpha}}{\alpha}\,{\rm tr}(d_{AB}) {\rm tr}(d_{CD})\,,
\end{eqnarray}
where we use the abbreviation ${\rm tr}(d_{AB})=\left( d_A^{ab} d_B^{ab} \right)$.

The other four contributions $I^{(2)}$, $I^{(4)}$, $I^{(6)}$, $I^{(8)}$ can be computed analogously. 
For example, using the abbreviation $(s d_{AB} s) = s_a s_c d_{A}^{ab} d_{B}^{bc} $, the expression for $I^{(2)}$ reads

\begin{equation}
\begin{aligned}
    I^{(2)}
= -  \frac{4\,j_1 (\alpha)}{\alpha} \ {\rm tr }(d_{AB})   \  {\rm tr } (d_{CD}) 
    &+ 2  j_2 (\alpha) \, (s d_{AB} s)  \  {\rm tr } (d_{CD}) +2 j_2 (\alpha)\, (s d_{CD} s)  \  {\rm tr } (d_{AB}) \, .  
\label{eq:GammaINT2}
\end{aligned}
\end{equation}

\subsection{An optimal estimator}

We now develop an estimator designed to optimally extract the 
connected part of the GW four-point correlation function. Our strategy follows the general philosophy of \cite{Allen:1997ad}, suitably modified to
the higher-correlator case. Throughout this section we keep the discussion as general as possible, without specifying a particular GW experiment.

Let the measured time-domain output of detector $a$ be
\begin{equation}
\Sigma(t) = s(t) + n(t) ,
\end{equation}
where $s(t)$ denotes the physical GW signal and $n(t)$ represents detector noise. 
We consider correlations among four such data streams $\Sigma(t)$ coming from independent detectors. 
The instrumental noise is assumed to be stationary, Gaussian, and uncorrelated between different detectors.
We work in the regime where the noise amplitude dominates over the signal, so that the variance of the estimator is governed by noise, while its expectation value is sourced entirely by the signal component $s(t)$. 
The GW signal measured by different detectors may exhibit non-Gaussian correlations, encoded in a non-vanishing four-point function $\langle s^4 \rangle$, which we assume to be stationary as discussed in Section~\ref{sec_trispectrum}.

\smallskip

To probe such correlations we construct an estimator built from three copies of the detector output. 
For a specific detector quartet $(a,b,c,d)$ we define
\begin{equation}
S_{abcd} =
\int_{-T/2}^{T/2} dt_1
\int_{-T/2}^{T/2} dt_2
\int_{-T/2}^{T/2} dt_3
\int_{-T/2}^{T/2} dt_4\,
\Sigma_a(t_1)\,\Sigma_b(t_2)\,\Sigma_c(t_3)\,\Sigma_c(t_4)\,
Q(t_2-t_1,t_3-t_1,t_4-t_1) ,
\label{def_sabc}
\end{equation}
where $T$ denotes the total observation time. 
The function $Q$ acts as a filter kernel and depends only on time differences, consistent with the stationarity assumption. It is taken to vanish for sufficiently large separations $|t_i-t_j|$. 
Our goal will be to determine the form of $Q$ that maximizes the response to the signal.

For the moment we focus on a single detector quartet $(a,b,c,d)$. 
The time series can be expressed in the frequency domain through
\begin{equation}
\Sigma(t) = \int_{-\infty}^{\infty} df \, e^{2\pi i f t}\,\tilde{\Sigma}(f) .
\end{equation}
Notice that frequencies run also along negative values. 
Substituting this representation into Eq.~\eqref{def_sabc} gives
\begin{equation}
S_{abcd} =
\int_{-\infty}^{\infty} df_1\,df_2\,df_3\,df_4 \,
\delta_T(f_1+f_2+f_3+f_4)\,
\tilde{\Sigma}_a(f_1)\,
\tilde{\Sigma}_b(f_2)\,
\tilde{\Sigma}_c(f_3)\,
\tilde{\Sigma}_d(f_4)\,
\tilde{Q}^*(f_2,f_3,f_4) ,
\end{equation}
where the finite-time delta function is defined as
\begin{equation}
\delta_T(f) \equiv
\int_{-T/2}^{T/2} dt\, e^{2\pi i f t},
\qquad
\delta_T(0)=T .
\end{equation}

In the noise-dominated limit the expectation value and variance of the statistic read
\begin{align}
\mu &= \langle S_{abcd} \rangle ,\\
\label{eq_defva}
\sigma^2 &= \langle S_{abcd}^2 \rangle - \mu^2 .
\end{align}
Because the detector noise is Gaussian, the mean $\mu$ receives contributions only from the GW signal, whereas the variance is determined by the noise. Since the signal is assumed much smaller than the noise, the $\mu^2$ term in Eq.~\eqref{eq_defva} can be neglected. 
The signal-to-noise ratio (SNR) of the estimator is therefore
\begin{equation}
\mathrm{SNR} = \frac{\mu}{\sigma} ,
\end{equation}
and the optimization problem reduces to finding the filter $Q$ that maximizes this quantity.

A straightforward calculation yields
\begin{equation}
\mu = T\,\kappa_{abcd}
\int_{-\infty}^{\infty} df_1\,df_2\,d f_3\,
{\cal T}_h(f_1,f_2,f_3)\,
\tilde{Q}^*(f_1,f_2,f_3) ,
\end{equation}
where $\kappa_{abcd}$ denotes the four-detector overlap reduction function and 
${\cal T}_h$ is the connected contribution to the GW trispectrum.
The frequencies satisfy the condition $f_1+f_2+f_3+f_4=0$.

Let $\sigma_a^2(f)$ denote the noise power spectral density for detector $a$. 
The noise correlator is then
\begin{equation}
\langle n_a(f_1)n_b(f_2)\rangle
=
\delta(f_1+f_2)\,
\delta_{ab}\,
\sigma_a^2(f_1) .
\end{equation}
Assuming Gaussian noise, the variance of the estimator becomes
\begin{equation}
\sigma^2 =
T
\int_{-\infty}^{\infty} df_1\,df_2\,d f_3\,
N_{abcd}(f_1,f_2,f_3)\,
|Q(f_1,f_2,f_3)|^2 ,
\end{equation}
with
\begin{equation}
N_{abc}(f_1,f_2,f_3)
=
\sigma_a^2(f_1)\,
\sigma_b^2(f_2)\,
\sigma_c^2(f_2)\,
\sigma_d^2(f_1+f_2+f_3)
+\mathrm{perms} .
\end{equation}

The resulting SNR can therefore be written as
\begin{equation}
\mathrm{SNR}
=
\sqrt{T}\,
\frac{
\kappa_{abcd}
\int df_1\,df_2\,d f_3\,
{\cal T}_h(f_1,f_2,f_3)\,
\tilde{Q}^*(f_1,f_2,f_3)
}{
\left[
\int df_1\,df_2\, d f_3\,
N_{abcd}(f_1,f_2,f_3)
|Q(f_1,f_2,f_3)|^2
\right]^{1/2}
}.
\end{equation}

To determine the optimal filter we employ the Wiener filtering method (see e.g. \cite{Allen:1997ad}). 
We introduce the inner product
\begin{equation}
(C,D)
\equiv
\int df_1\,df_2\,d f_3\,
C(f_1,f_2,f_3)
D^*(f_1,f_2,f_3)
N_{abcd}(f_1,f_2,f_3) ,
\end{equation}
which allows us to rewrite the SNR as
\begin{equation}
\mathrm{SNR}
=
\sqrt{T}\,
\frac{
\left(
\kappa_{abcd} {\cal T}_h / N_{abcd},
Q
\right)
}{
\left[(Q,Q)\right]^{1/2}
}.
\end{equation}

Maximization with respect to $Q$ leads to the optimal choice
\begin{equation}
\label{eq_opf}
Q(f_1,f_2,f_3)
=
\frac{
\kappa_{abcd}\,
{\cal T}_h(f_1,f_2,f_3)
}{
N_{abcd}(f_1,f_2,f_3)
}.
\end{equation}
Substituting this expression back into the SNR yields the maximal achievable value
\begin{equation}
\mathrm{SNR}_{\rm max}
=
\sqrt{T}
\left[
\int df_1\,df_2 \,d f_3
\frac{
\big(\kappa_{abcd} {\cal T}_h(f_1,f_2,f_3)\big)^2
}{
N_{abcd}(f_1,f_2,f_3)
}
\right]^{1/2}.
\end{equation}

Further simplification arises when the noise spectrum is approximately frequency independent,
\begin{equation}
\langle n_a(f_1)n_b(f_2)\rangle
=
\delta(f_1+f_2)\,
\delta_{ab}\,
\sigma_a^2 ,
\end{equation}
which implies
\begin{equation}
N_{abcd}
=
4\,\sigma_a^2\,\sigma_b^2\,\sigma_c^2\,\sigma_d^2 .
\end{equation}
In this case the maximal signal-to-noise ratio becomes
\begin{equation}
\mathrm{SNR}_{\mathrm{max}}
=
\sqrt{\frac{T}{4}}
\left[
\int df_1\,df_2\,d f_3
\frac{
\big(\kappa_{abcd} {\cal T}_h(f_1,f_2,f_3)\big)^2
}{
\sigma_a^2 \sigma_b^2 \sigma_c^2 \sigma_d^2
}
\right]^{1/2}.
\label{eq_opesi}
\end{equation}

The result above applies to a single detector quadruplet. 
In experiments where multiple independent quadruplets can be formed, such as pulsar timing arrays, the total signal-to-noise ratio can be enhanced by summing the contributions from all available combinations.

\subsection*{ Acknowledgments}

It is a pleasure to thank Jinn-Ouk Gong, Maria Mylova, and Misao Sasaki for discussions on related subjects. GT is partially funded by the STFC grants ST/T000813/1 and ST/X000648/1.

\begin{appendix}
\section{The time integral of Eq.~\eqref{def_tintA}}
\label{app_time}
We present here the derivation of the late-time limit result of the integral in eq.~\eqref{def_tintA} in radiation dominated era.\\
The integral to compute is
\begin{equation}
    J_\tau(\tau)\,=\,\prod_{i=1}^4\,\int d \tau_i\,\frac{
g_{k_i} (\tau,\tau_i)}{a(\tau_i)}
\end{equation}
where
\begin{equation}
g_{k_i}(\tau,\tau_i)
=
\frac{i}{2k_i}
\left(
e^{-ik_i(\tau-\tau_i)}-e^{ik_i(\tau-\tau_i)}
\right) .
\end{equation}
The result of the full integral is
\begin{equation}
\begin{aligned}
&J_\tau(\tau)=\frac{1}{16k_1k_2k_3k_4}
\Big[
-\mathrm{Ci}(k_1 \tau_R)\,\sin(k_1\tau)
+\mathrm{Ci}(k_1 \tau)\,\sin(k_1\tau)
+\cos(k_1\tau)\big(\mathrm{Si}(k_1\tau_R)-\mathrm{Si}(k_1\tau)\big)
\Big] \\
&\quad\times
\Big[
-\mathrm{Ci}(k_2 \tau_R)\,\sin(k_2\tau)
+\mathrm{Ci}(k_2 \tau)\,\sin(k_2\tau)
+\cos(k_2\tau)\big(\mathrm{Si}(k_2\tau_R)-\mathrm{Si}(k_2\tau)\big)
\Big] \\
&\quad\times
\Big[
-\mathrm{Ci}(k_3 \tau_R)\,\sin(k_3\tau)
+\mathrm{Ci}(k_3 \tau)\,\sin(k_3\tau)
+\cos(k_3\tau)\big(\mathrm{Si}(k_3\tau_R)-\mathrm{Si}(k_3\tau)\big)
\Big] \\
&\quad\times
\Big[
-\mathrm{Ci}(k_4 \tau_R)\,\sin(k_4\tau)
+\mathrm{Ci}(k_4 \tau)\,\sin(k_4\tau)
+\cos(k_4\tau)\big(\mathrm{Si}(k_4\tau_R)-\mathrm{Si}(k_4\tau)\big)
\Big].
\end{aligned}
\end{equation}
Using the asymptotic behaviour $\mathrm{Ci}(k\tau)\xrightarrow{\tau\to\infty}0$, $\mathrm{Si}(k\tau)\xrightarrow{\tau\to\infty}\displaystyle\frac{\pi}{2}$, expanding the $\cos$ and $\sin$ functions in exponential form and renaming $\mathcal{C}(k_i)=(16k_1k_2k_3k_4)^{-1}$ we find
\begin{equation}
\begin{aligned}
&J_\tau(\tau)= \frac{{C}(k_i)}{16} [ \pi^4e^{-i (k_1+k_2+k_3+k_4)\tau}
+\pi^4e^{2 i k_1 \tau-i (k_1+k_2+k_3+k_4)\tau}
+\pi^4e^{2 i k_2 \tau-i (k_1+k_2+k_3+k_4)\tau}\\
&\quad +\cdots\,(1938)\,\cdots \,] \,+\\
&\quad
+e^{2 i k_1 \tau+2 i k_3 \tau+2 i k_4 \tau-i (k_1+k_2+k_3+k_4)\tau}\,
\text{Si}[k_1\tau_R]\,
\text{Si}[k_2\tau_R]\,
\text{Si}[k_3\tau_R]\,
\text{Si}[k_4\tau_R] \\
&\quad
+e^{2 i k_2 \tau+2 i k_3 \tau+2 i k_4 \tau-i (k_1+k_2+k_3+k_4)\tau}\,
\text{Si}[k_1\tau_R]\,
\text{Si}[k_2\tau_R]\,
\text{Si}[k_3\tau_R]\,
\text{Si}[k_4\tau_R] \\
&\quad
+e^{2 i k_1 \tau+2 i k_2 \tau+2 i k_4 \tau-i (k_1+k_2+k_3+k_4)\tau}\,
\text{Si}[k_1\tau_R]\,
\text{Si}[k_2\tau_R]\,
\text{Si}[k_3\tau_R]\,
\text{Si}[k_4\tau_R] \\
&\quad
+e^{2 i k_1 \tau+2 i k_2 \tau+2 i k_3 \tau-i (k_1+k_2+k_3+k_4)\tau}\,
\text{Si}[k_1\tau_R]\,
\text{Si}[k_2\tau_R]\,
\text{Si}[k_3\tau_R]\,
\text{Si}[k_4\tau_R] \, .
\end{aligned}
\end{equation}
Without conditions, taking the average over rapid oscillations leads to a vanishing result.\\
By imposing that $s_i=+1$ for all $i$, and $k_4=k_1+k_3-k_2$, we get
\begin{align*}
J_\tau(\tau) &= {C}(k_i) \Bigg\{
\mathrm{Ci}(k_3\tau_R)\Bigg[
\frac{\mathrm{Ci}(k_2\tau_R)}{2\pi}\Big(\pi-2\,\mathrm{Si}(k_1\tau_R)\Big)
-\frac{\mathrm{Ci}(k_1\tau_R)}{2\pi}\Big(\pi-2\,\mathrm{Si}(k_2\tau_R)\Big)
\Bigg]\\
&+\frac{1}{8\pi^{2}}
\Big(\pi-2\,\mathrm{Si}(k_1\tau_R)\Big)\Big(\pi-2\,\mathrm{Si}(k_2\tau_R)\Big)
\nonumber\\
&-\frac{1}{4\pi}
\Big(\pi-2\,\mathrm{Si}(k_1\tau_R)\Big)\Big(\pi-2\,\mathrm{Si}(k_2\tau_R)\Big)\,
\mathrm{Si}(k_3\tau_R)
\nonumber\\[1.5ex]
&+\mathrm{Ci}(k_1\tau_R)\,\mathrm{Ci}(k_2\tau_R)
\Big(\frac{\pi^{2}}{2}-\pi\,\mathrm{Si}(k_3\tau_R)\Big)
\nonumber+\mathrm{Ci}\,\!\Big((k_1-k_2+k_3)\tau_R\Big) \\
&\times \Bigg\{
\mathrm{Ci}(k_3\tau_R)\Bigg[
2\,\mathrm{Ci}(k_1\tau_R)\,\mathrm{Ci}(k_2\tau_R)
+\frac{1}{2}\Big(\pi-2\,\mathrm{Si}(k_1\tau_R)\Big)\Big(\pi-2\,\mathrm{Si}(k_2\tau_R)\Big)
\Bigg]
\nonumber\\[1.5ex]
&
+\mathrm{Ci}(k_2\tau_R)\Big(\pi-2\,\mathrm{Si}(k_1\tau_R)\Big)
\Big(-\frac{\pi}{2}+\mathrm{Si}(k_3\tau_R)\Big)
\nonumber\\[1.5ex]
&
+\mathrm{Ci}(k_1\tau_R)\Bigg[
\frac{1}{2\pi}\Big(\pi-2\,\mathrm{Si}(k_2\tau_R)\Big)
+\Big(-\pi+2\,\mathrm{Si}(k_2\tau_R)\Big)\mathrm{Si}(k_3\tau_R)
\Bigg]
\Bigg\}
\nonumber\\[1ex]
&+\Bigg\{
\mathrm{Ci}(k_3\tau_R)\Big[
\mathrm{Ci}(k_2\tau_R)\Big(-\pi+2\,\mathrm{Si}(k_1\tau_R)\Big)
+\mathrm{Ci}(k_1\tau_R)\Big(\pi-2\,\mathrm{Si}(k_2\tau_R)\Big)
\Big]
\nonumber\\[1ex]
&
-\frac{1}{4\pi}\Big(\pi-2\,\mathrm{Si}(k_1\tau_R)\Big)\Big(\pi-2\,\mathrm{Si}(k_2\tau_R)\Big)\\
&+\frac{1}{2}\Big(\pi-2\,\mathrm{Si}(k_1\tau_R)\Big)\Big(\pi-2\,\mathrm{Si}(k_2\tau_R)\Big)\mathrm{Si}(k_3\tau_R)
\nonumber\\[1ex]
&
+\mathrm{Ci}(k_1\tau_R)\,\mathrm{Ci}(k_2\tau_R)\Big(-\pi+2\,\mathrm{Si}(k_3\tau_R)\Big)
\Bigg\}\;
\mathrm{Si}\,\!\Big((k_1-k_2+k_3)\tau_R\Big)\Bigg\}.
\end{align*}
By also imposing $k_2 =k_1$ and $k_3 = k_1$, corresponding to the configuration in the right panel of fig. \ref{fig_quad},  we get our familiar result of eq. \eqref{val_itau2}
\begin{equation}
\frac{1}{8k_1^4H_0^4\,\Omega_{\rm RD}^2}\left[\mathrm{Ci}(k_1\tau_R)^2+\left(\frac{\pi}{2}-\mathrm{Si}(k_1\tau_R)\right)^2\right]^2
\end{equation}
where we have restored the $H_0^4\,\Omega_{\rm RD}^2$ at denominator, since in the computation we have simply used $a(\tau)=\tau$.
\end{appendix}

{\small
\providecommand{\href}[2]{#2}\begingroup\raggedright\endgroup

}

\end{document}